\documentclass[useAMS,usenatbib,apjstyle,twocolumn]{aastex61}
\usepackage{epsfig, bm, times, aas_macros}
\usepackage[english]{babel}
\usepackage{amsmath}
\usepackage{amssymb}
\usepackage{graphicx}
\usepackage{xcolor}
\newcommand{\trifig}[3]{
\includegraphics[width=0.3\textwidth]{#1}\hspace{\stretch{1}}%
\includegraphics[width=0.3\textwidth]{#2}\hspace{\stretch{1}}%
\includegraphics[width=0.3\textwidth]{#3}%
}
\newcommand{\twofig}[2]{
\includegraphics[width=0.47\textwidth]{#1}\hspace{\stretch{1}}%
\includegraphics[width=0.47\textwidth]{#2}%
}
\newcommand{\twofigL}[2]{
  \begin{center}
    \includegraphics[width=0.9\textwidth]{#1}\\%\hspace{\stretch{1}}%
    \includegraphics[width=0.9\textwidth]{#2}%
  \end{center}
}

\newcommand{\seclab}{Section}

\newcommand{\figlab}{Fig}
\newcommand{\equlab}{Eq}
\newcommand{\secref}[1]{\seclab{} \ref{#1}}

\newcommand{\figref}[1]{\figlab{.} \ref{#1}}
\newcommand{\figrefs}[1]{\figlab{s.} \ref{#1}}
\newcommand{\eqr}[1]{\equlab{.} \eqref{#1}}

\newcommand{\tent}[1]{\times 10^{#1}}

\newcommand{\x}{y}
\newcommand{\y}{x}
\newcommand{\uy}{U_{\rm \y}}

\newcommand{\appint}{Finally, in the appendix we show snapshots in 3D
  in order to provide a global view of the evolution of the
  perturbation that we describe below.}

\newcommand{\appdo}{%
\appendix

Here we show snapshots in 3D in order to provide a global view of the
evolution of the perturbation, \figref{fig:3d}. The ``outside'' surfaces
are colour coded according to the temperature, while the ``inside'' is
colour coded according to the nuclear burning yield.

\begin{figure*}[!h]
  \includegraphics[width=0.49\textwidth]{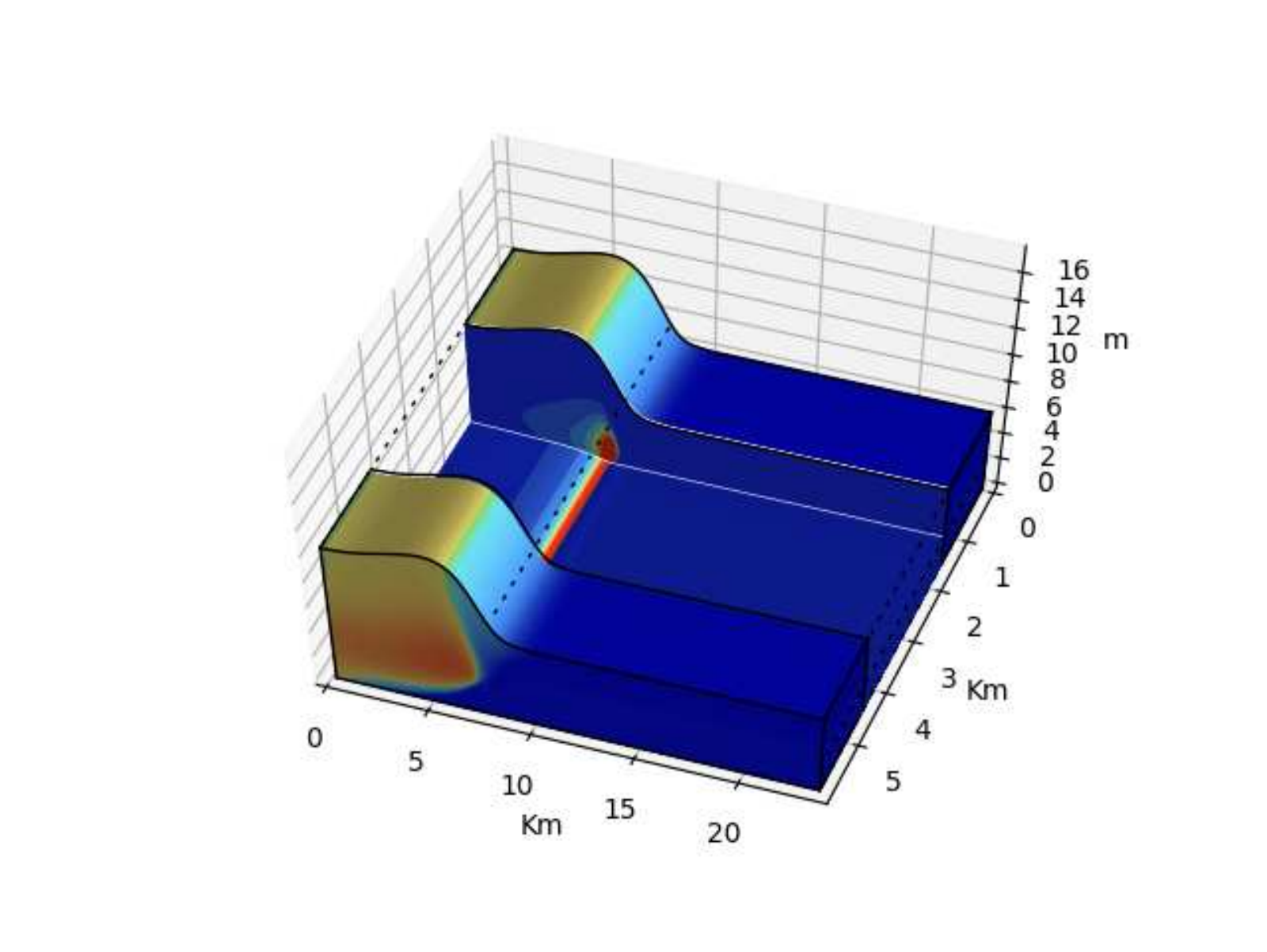}\hspace{\stretch{1}}%
  \includegraphics[width=0.49\textwidth]{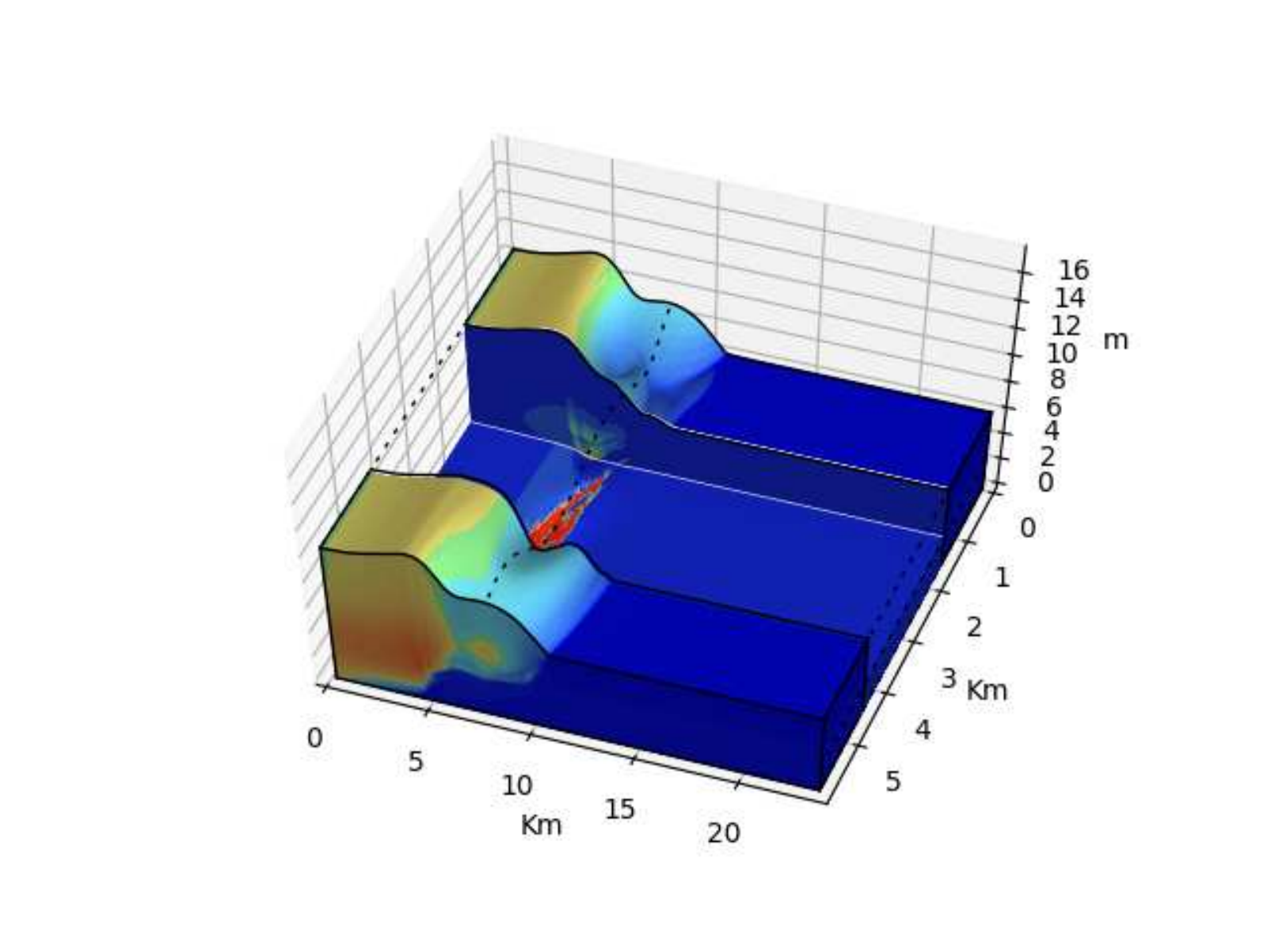}\\%
  \includegraphics[width=0.49\textwidth]{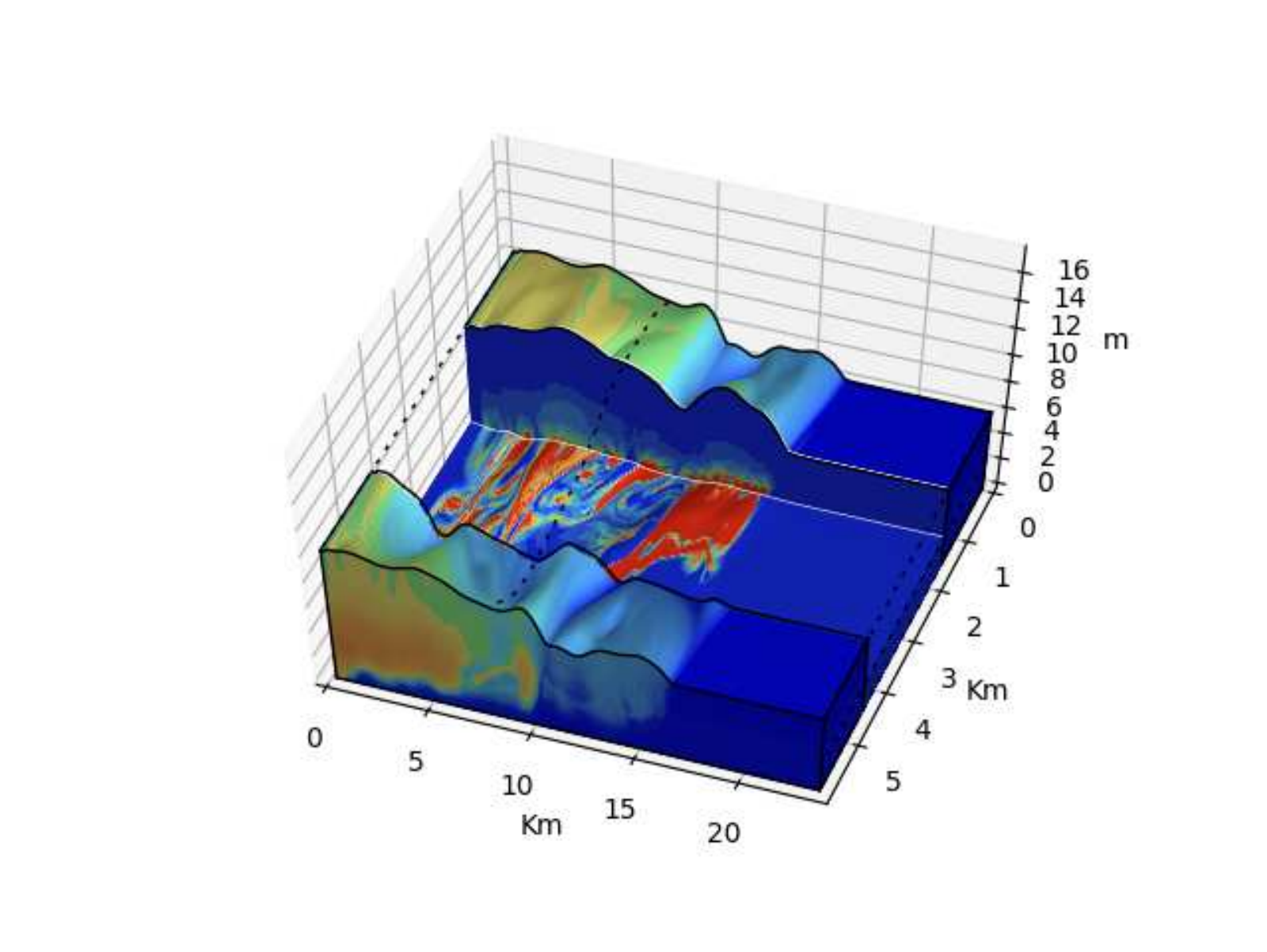}\hspace{\stretch{1}}%
  \includegraphics[width=0.49\textwidth]{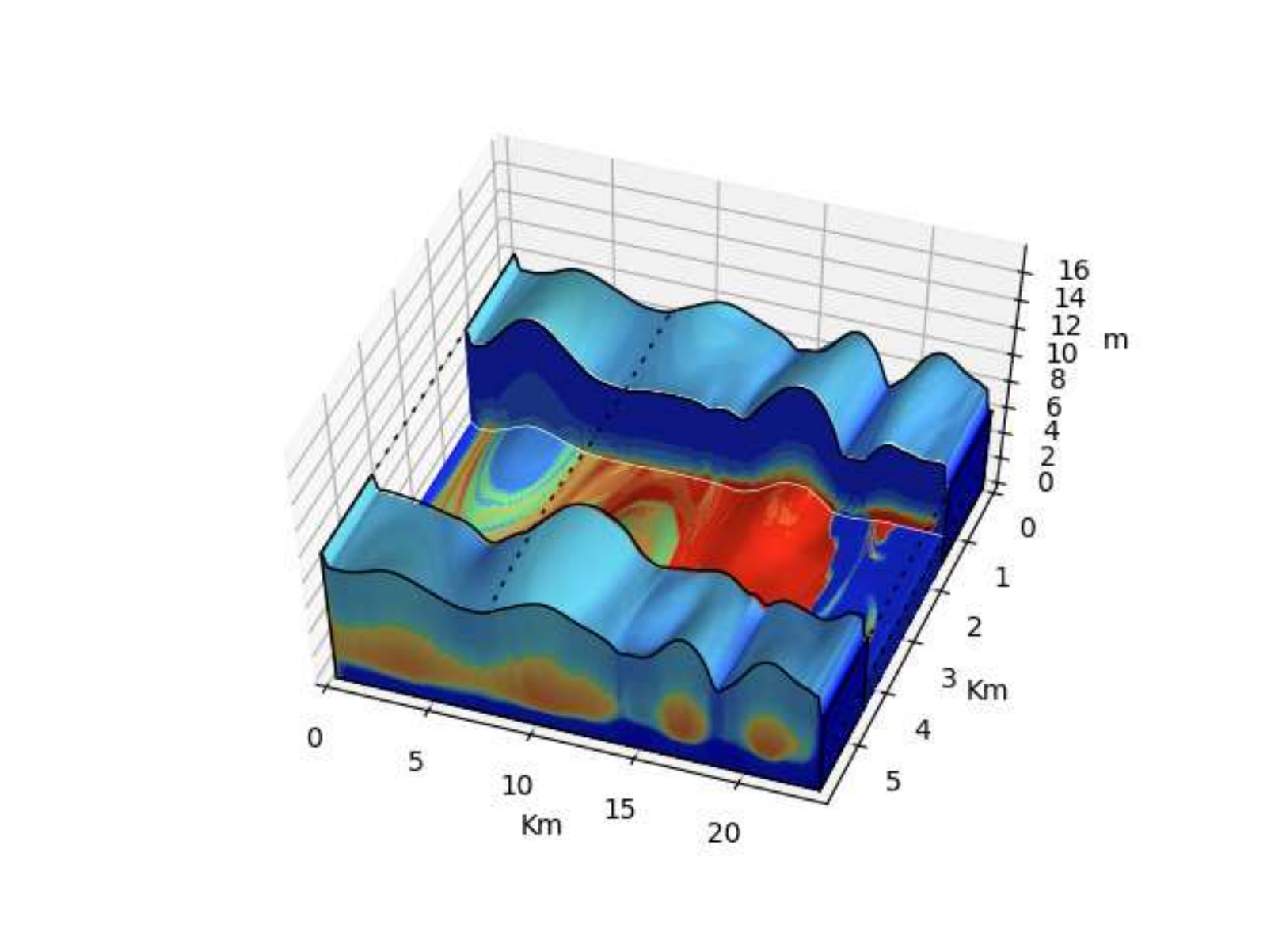}%
  \caption{3D rendering of the evolution of the flame front. The
    surfaces ``outside'' are colour coded according to the
    temperature, while the contours in the ``inside'' follow the
    nuclear burning yield. Note that for visibility here the aspect
    ratio between $\y$ and $\x$ is not 1.}
  \label{fig:3d}
\end{figure*}
}

\shorttitle{The 3D instability of Type I burst flames}
\shortauthors{Cavecchi \& Spitkovsky}

\begin{document}

\title{Three-dimensional instability of flame fronts in type I X-ray bursts}

\author[0000-0002-6447-3603]{Yuri Cavecchi}
\affiliation{Department of Astrophysical Sciences, Princeton University,
  Peyton Hall, Princeton, NJ 08544, USA}
\affiliation{Mathematical Sciences and STAG Research Centre,
  University of Southampton, SO17 1BJ, UK}

\author{Anatoly Spitkovsky}
\affiliation{Department of Astrophysical Sciences, Princeton University,
  Peyton Hall, Princeton, NJ 08544, USA}

\correspondingauthor{Yuri Cavecchi}
\email{cavecchi@astro.princeton.edu,y.cavecchi@soton.ac.uk}

\label{firstpage}

\begin{abstract}
% the abstract
  We present the first realistic 3D simulations of flame front
  instabilities during type I X-ray bursts. The unperturbed front is
  characterised by the balance between the pressure gradient and the
  Coriolis force of a spinning neutron star ($\nu = 450$ Hz in our
  case). This balance leads to a fast horizontal velocity field
  parallel to the flame front. This flow is strongly sheared in the
  vertical direction. When we perturb the front an instability quickly
  corrugates the front. We identify this instability as the baroclinic
  instability. Most importantly, the flame is not disrupted by the
  instability and there are two major consequences: the overall flame
  propagation speed is $\sim 10$ times faster than in the unperturbed
  case and distinct flame vortices appear. The speedup is due to the
  corrugation of the front and the dynamics of the vortices. These
  vortices may also be linked to the oscillations observed in the
  lightcurves of the bursts.
\end{abstract}

% the keywords

\section{Introduction}

The type I X-ray bursts are thermonuclear explosions that burn the
material that neutron stars accrete from their companion in low mass
X--ray binaries.  The matter is stripped from the companion via
Roche-Lobe overflow and forms an accretion disk around the neutron
star. It eventually reaches the star and spreads across its surface.
With time, the fresh material is compressed by newly incoming one, and
it sinks to deeper, higher pressure layers. In this process the fuel,
which is rich in hydrogen and helium, undergoes nuclear
burning. Depending on the composition and the mass accretion rate, the
heating from the nuclear burning may become uncompensated by the
cooling processes, and the temperature rapidly increases leading to a
thermonuclear runaway: the ignition of type I bursts
\citep{art-1981-fuji-han-miy, rev-1998-bild,
  rev-2003-2006-stro-bild-book, rev-2017-gal-keek}.

One important aspect of the bursts is flame propagation: the fuel is
assumed to cover the whole star, but synchronous ignition is unlikely
\citep{art-1982-shara}, and the flame needs to propagate from the
first ignition site to engulf the star. The most likely ignition site
is supposed to be the equator \citep{art-2002-spit-levin-ush}.  The
flame propagation needs to be fast enough to be compatible with the
$\sim 1$ s rise time of the burst lightcurves and yet cannot be a
detonation, since the necessary conditions for detonations are not
reachable \citep{art-1982-fryx-woos-b,art-2011-malo-etal}. This poses
stringent constraints on the flame spreading mechanism. Furthermore,
the flame propagation (and the subsequent cooling of the surface)
should explain the temperature distribution across the star surface
and thus the observed features of the observed lightcurves.

The details of burst flames have been explored in a series of papers:
\citet{art-2011-malo-etal} and \citet{art-2015-zing-etal} explored the
onset of nuclear burning ignition and the convection driven by nuclear
heating. \citet{art-2002-spit-levin-ush} presented shallow water
calculation suggesting the Coriolis force should be important for
ignition and front propagation.  \citet{art-2013-cavecchi-etal,
  art-2015-cavecchi-etal, art-2016-cavecchi-etal} explored the
vertical structure of the front and its spreading across the surface.
The picture emerging from these works is the following: the hot
burning fluid expands vertically, creating a horizontal pressure
gradient which pushes the hot fluid towards the cold one. The Coriolis
force balances the pressure gradient (geostrophic balance),
generating a horizontal flow parallel to the flame front. This flow
follows the thermal wind profile \citep[ex.][]{book-1987-Pedlo}
showing a strong vertical shear. The hot fluid is thus being confined
within a few Rossby radii. The Rossby radius is the radius over which
the Coriolis force is capable of deflecting the horizontal motion,
$R_{\rm Ro} ~= \sqrt{g H} / 2 \Omega$, where $\Omega$ is the angular
velocity of the star, $g$ the gravitational acceleration and $H$ the
scale height of the fluid.

The flame propagation is driven mostly by conduction across the
hot-cold fluid interface at the front. The latter is almost parallel
to the horizontal, since the Rossby radius is of order $5\tent{4}$ cm
while the layer height is of order $10^2$ cm, and the large conducting
surface guarantees high speed in good agreement with observations.
The tension provided by a sufficiently strong magnetic field, $\sim
10^{8}$ G, will compete with the Coriolis force, reducing the thermal
wind and increasing the confinement length, thus making the front even
more horizontal, increasing the conducting surface and speeding up the
flame \citep{art-2016-cavecchi-etal}.

One important question left unanswered by these studies is the
stability of the flame front. The thermal wind can be highly unstable
to the baroclinic instability \citep[see ex.][]{book-1987-Pedlo,
  book-2006-vallis}. In a nutshell, baroclinic instability is a kind
of \emph{quasi horizontal} convection which taps the available
gravitational potential energy when the surfaces of constant pressure
and the surfaces of constant density are not coincident. Baroclinic
instability has been previously discussed within the burst framework
by e.g. \citet{art-1988-fuji} and \citet{art-2000-cum-bild}. These
authors addressed angular momentum and mixing in differentially
rotating atmospheres, in particular following the expansion of the
burning fluid. \citet{art-2013-cavecchi-etal} were the first to
identify this instability along the fronts of burst flames. However,
their simulations were 2.5D, assuming symmetry along the $y$ direction
(parallel to the flame front), and the instability could only develop
on the vertical plane, bringing cold unburnt fluid from the top of the
flame front down to the highest burning region, thus helping
propagation.

In this paper we study the full development of front instabilities in
3D. The description of the numerical setup is given in section
\secref{sec:numset}. In \secref{sec:res} we describe the development
of the instability and discuss the results in \secref{sec:discu}.

\section{Numerical setup}
\label{sec:numset}

\begin{figure*}
  \twofig{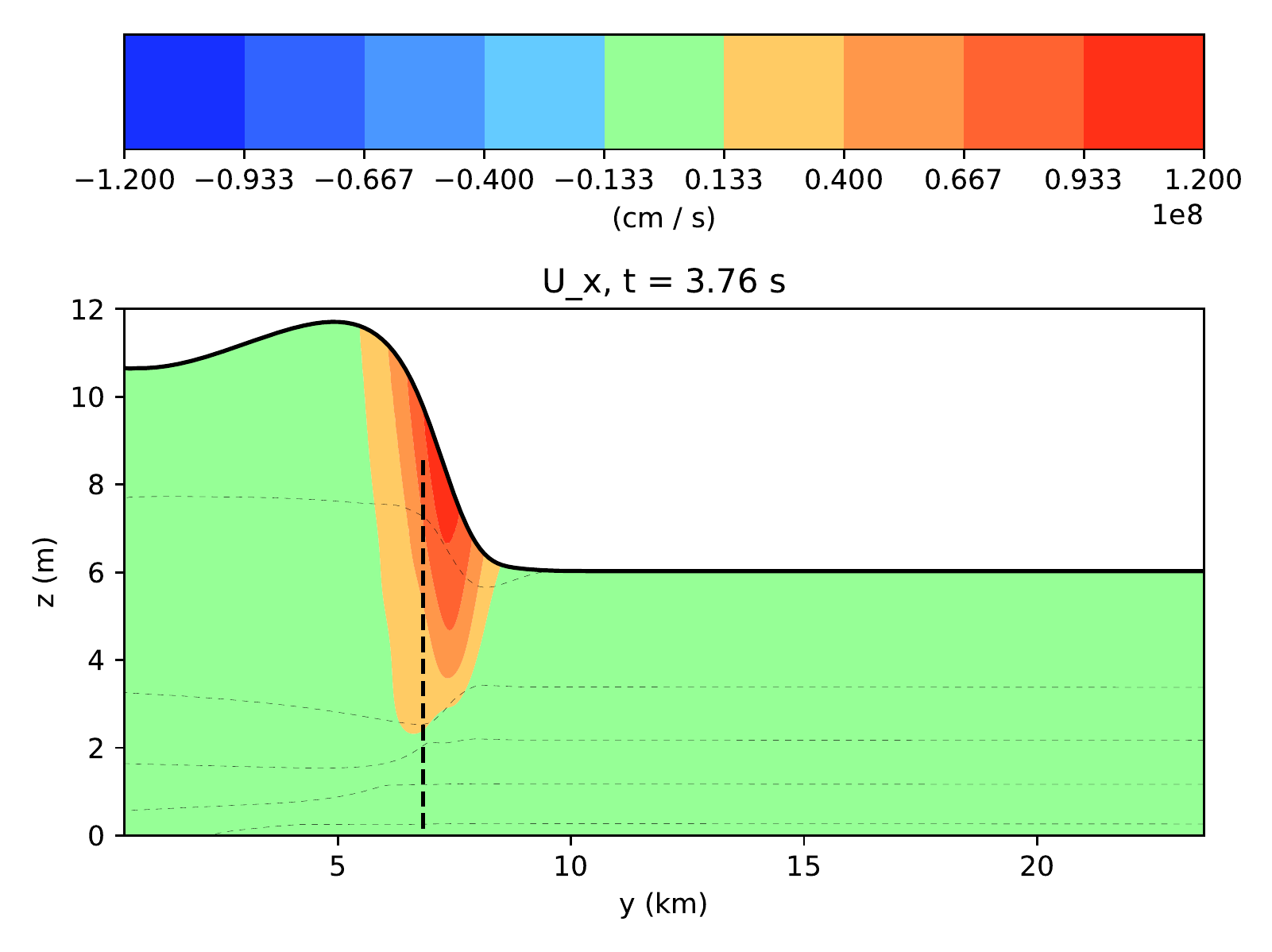}{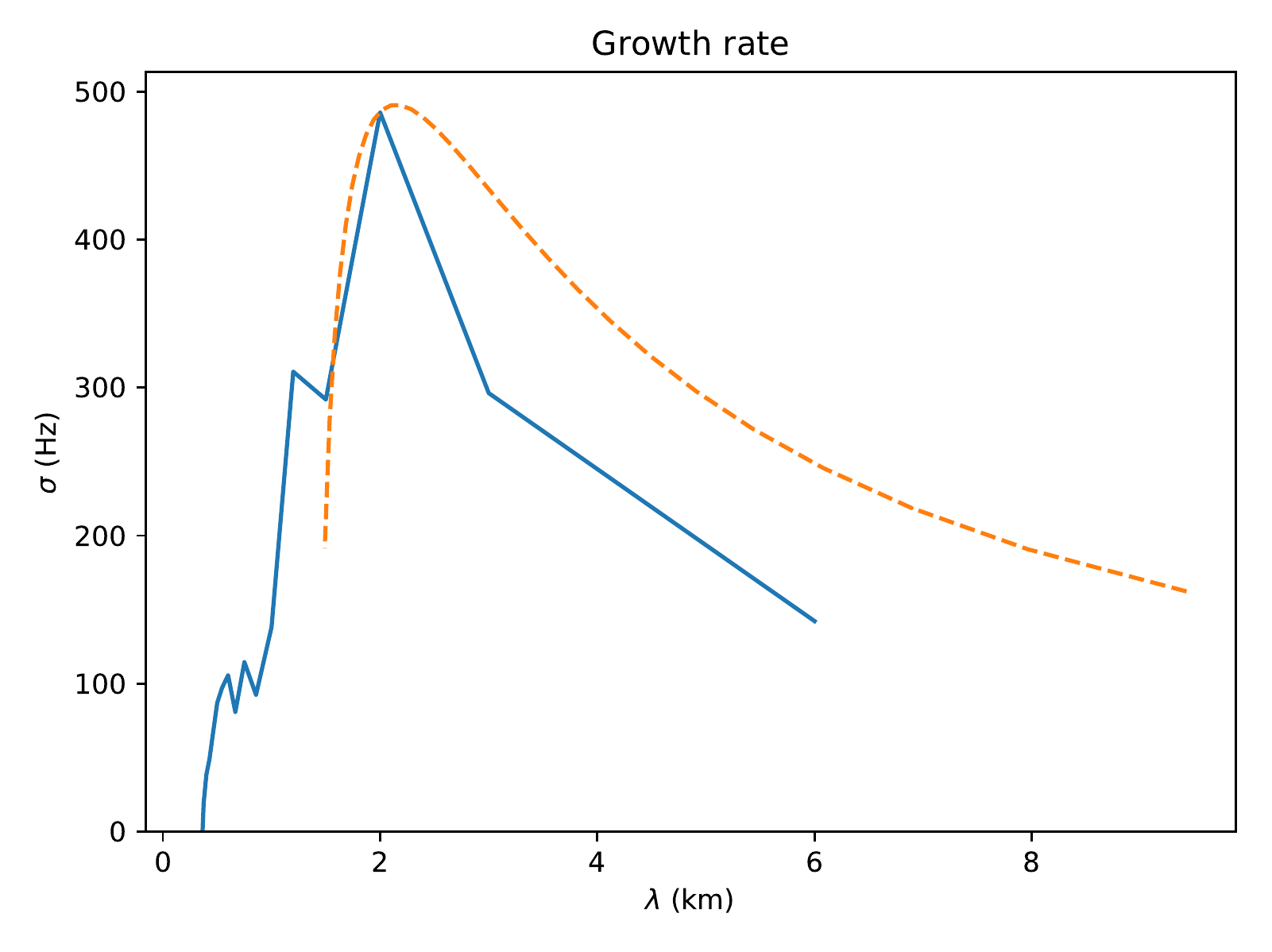}
  \caption{\textbf{Left panel}: the initial velocity profile and the
    location of the perturbation. Dotted lines are contours of
    density, increasing downwards. \textbf{Right panel}: the growth
    rate of the instability at the early stages. The dashed line is
    \eqr{equ:eady} with parameters: $R_{\rm Ro} = 5\tent{4}$ cm,
    $U = 10^8$ cm s$^{-1}$, $L = 3 R_{\rm Ro}$.}
  \label{fig:uyvertprofs}
\end{figure*}

In order to study the instability we first ignite and let the flame
propagate in a 2.5D simulation as in \citet{art-2013-cavecchi-etal,
  art-2015-cavecchi-etal, art-2016-cavecchi-etal}, where the $\x$
direction has to be thought of as the equator to pole direction and the
$z$ direction is the radial direction. We then extend the simulation in
the third dimension ($\y$, west to east), perturb the velocity field and
let the instability set in.

The code we used is described in \citet{art-2012-braith-cavecchi} and
\citet{art-2013-cavecchi-etal, art-2015-cavecchi-etal,
  art-2016-cavecchi-etal}, and specific details can be found there. For
this paper we used only the hydrodynamical modules, switching the
magnetic field off. The code is a finite differences, explicit time
stepping code, using a 3-step Runge-Kutta scheme designed to save
memory impact \citep{art-1980-will}. Derivatives, interpolations and
integrals use six point stencils on a staggered grid, that reach
spectral-like accuracy \citep{art-1992-lele}. Velocities in the three
directions are evaluated on the cell faces, state variables like
density, temperature and mass fraction of the different species are
evaluated at the cell centres.

The equations solved assume hydrostatic equilibrium in the vertical
direction and use pressure as the vertical coordinate. We simulate a
box with reflecting boundary conditions in the $\x$ direction and
periodic boundaries in the $\y$ direction. Vertically, we use
symmetrical conditions for horizontal velocities and the state
variables like temperature. We further impose cooling losses from the
top, approximating radiative conditions at the top of the atmosphere
\citep[see Eq. 35 in][]{art-2013-cavecchi-etal}. The domain extends
for $24$ km along the $\x$ direction (corresponding to a $\sim 15$ km
radius star), for $6$ km in the $\y$ direction and is initially $6$ m
high.  The number of grid points in the three directions is $n_\y = 96$,
$n_\x = 768$ and $n_z = 168$.  The initial temperature profile is as in
\citet{art-2016-cavecchi-etal}:

\begin{equation}
T = T_0 + \frac{\delta T }{1 + \exp[(\x - \x_0)/\delta \x]}
\end{equation}
Using $T_0=2\tent{8}$ K, $\delta T=2.81\tent{8}$ K, $\x_0 = 9\tent{4}$
cm and $\delta \x=3.6\tent{4}$ cm.

\begin{figure*}
  \twofig{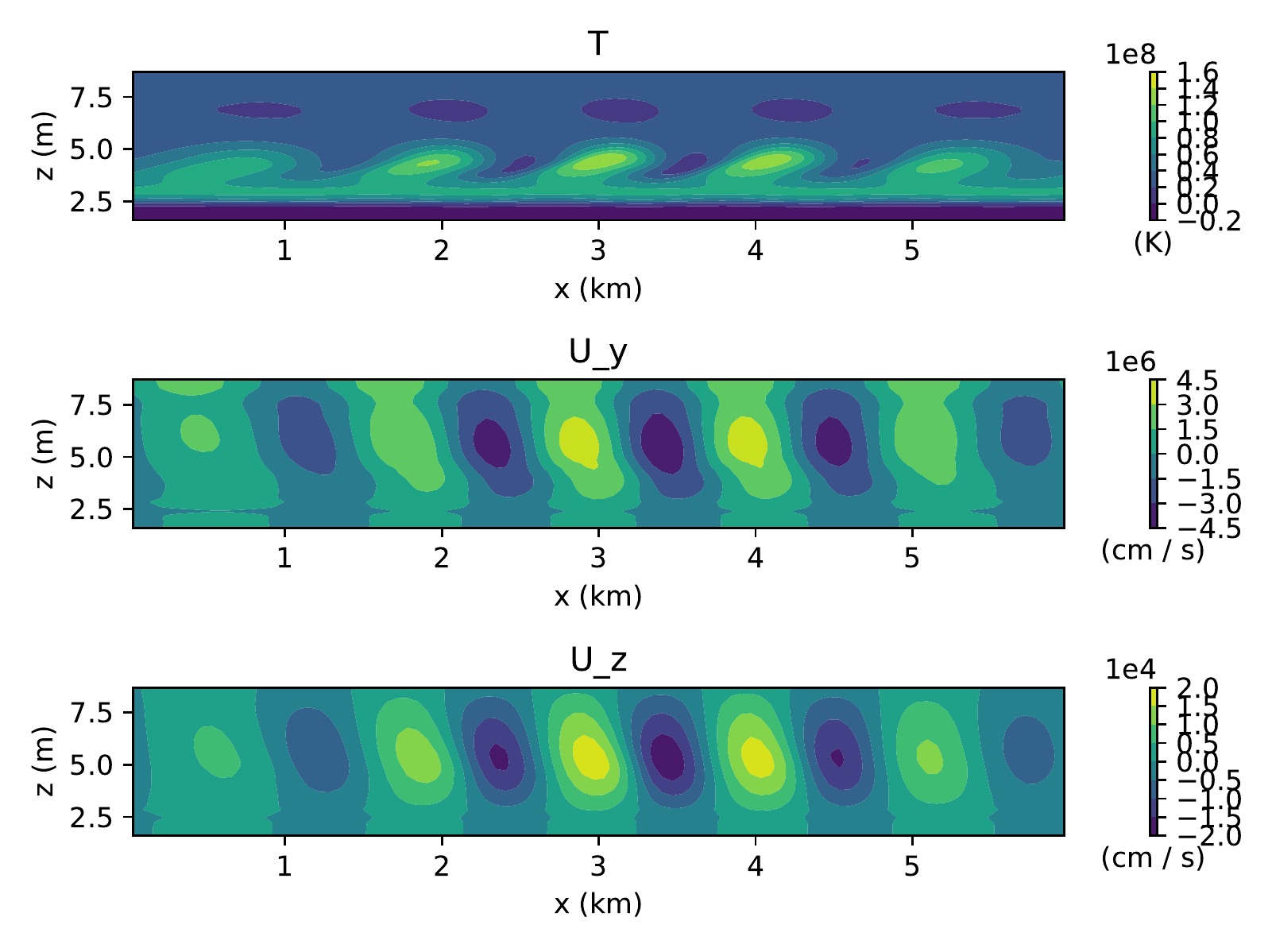}{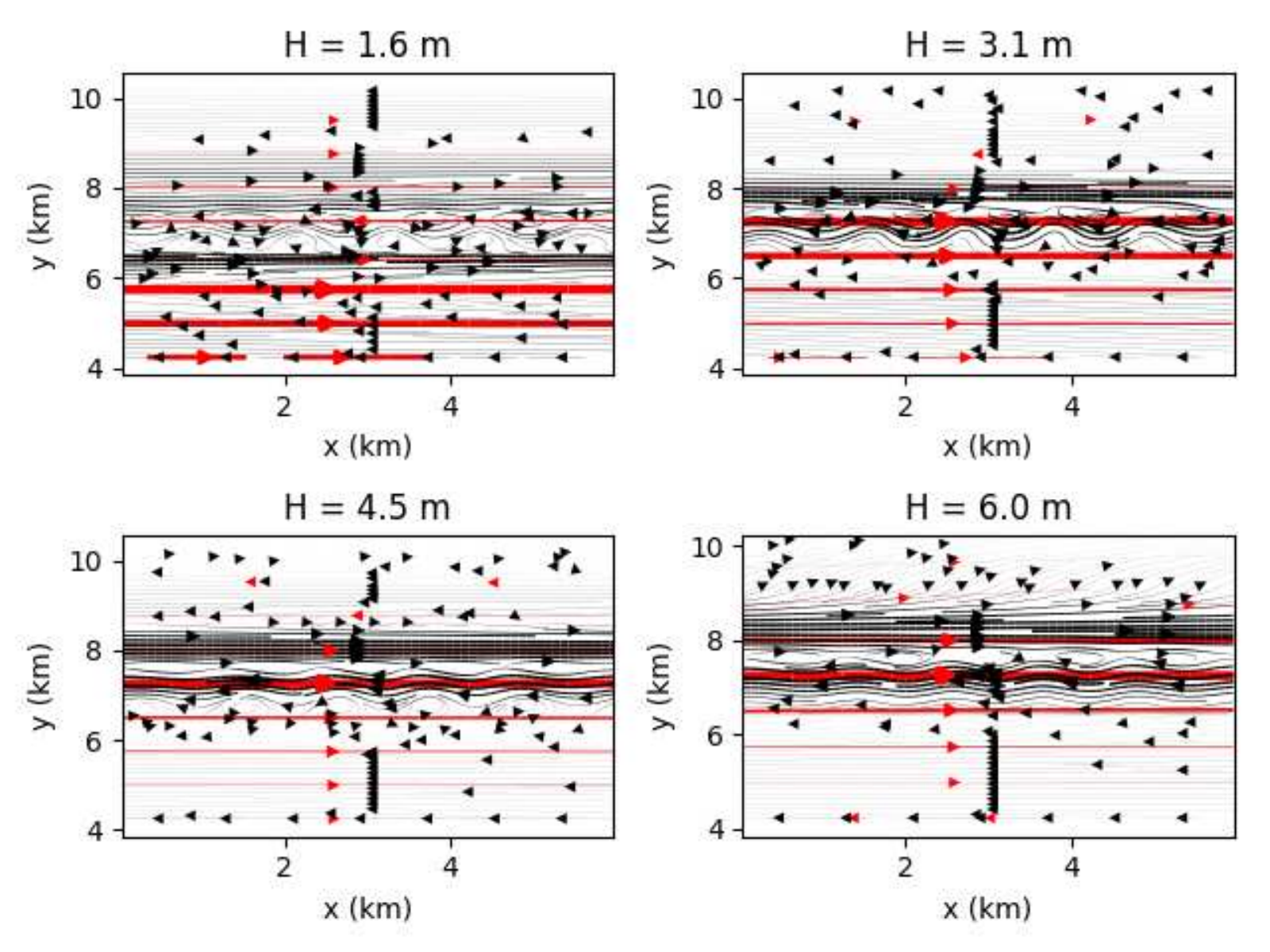}
  \caption{\textbf{Left panel}: The vertical profile of temperature,
    meridional and vertical velocities as a function of azimuth in the
    early stages of the instability. \textbf{Right panel}: horizontal
    slices showing streamlines of the background flow (red) and of the
    perturbations (black). The thickness of the streamlines is
    proportional to the speed, but for visibility each panel uses a
    different scale. The typical profile of the velocity contours
    \emph{leaning} against the background flow is evident especially
    in the left panel. The perturbation is bringing hot fluid up and
    northwards and cold fluid down and southwards.}
  \label{fig:instabgrow}
\end{figure*}

Differently from our previous simulations we add a layer of hydrogen
at the top and thicken the layer of carbon at the bottom. Our reaction
network only burns helium into carbon through triple alpha reactions,
so that hydrogen is not involved in the flame propagation. However, it
provides a layer of light fluid on top of the burning fluid, emulating
the role of newly accreted matter and providing space for the
instability to develop into. A similar role is played by the carbon
layer, which also emulates the ashes of previous bursts. The
initial profiles of the mass fraction $X$ of hydrogen and $Y$ of helium
are initialised as a function of pressure $P$:

\begin{equation}
  X = A \cdot B, \;\;\;\;\; Y = (1 - A) \cdot B
\end{equation}
with
\begin{align*}
  A =&     \frac{1}{1 + \exp[(\sigma -  3.2\tent{-2})/1.6\tent{-2}]}\\
  B =& 1 - \frac{1}{1 + \exp[(3.16\tent{-1} - \sigma)/1.6\tent{-2}]}
\end{align*}
and $\sigma = (P - P_{\rm top}) / (P_{\rm bot} - P_{\rm top})$, it is
$0$ at the top and $1$ at the bottom. $P_{\rm top} = 5\tent{21}$ erg
cm$^{-3}$ and $P_{\rm bot} = 1.6\tent{23}$ erg cm$^{-3}$. The carbon
fraction is $Z = 1 - X - Y$.

To ensure numerical stability we have added two regions of higher
viscosity at the equatorial and polar walls and at the bottom of the
carbon layer.  The flame front perturbation takes place far enough in
the simulation domain that these regions do not affect the dynamics of
the instability. Finally, in order to be able to see the effects of
cooling more quickly we also enhance the cooling from the top by a
factor of $5$.

When the flame has reached steady propagation, we extend the domain in
the $\y$ direction and perturb the $\uy$ velocity field in the
following way. We identify a location $(\x_0, z_0)$ near the highest
burning and extract the value of $\uy^0$ there. We then subtract the
value $\uy^0 f(x, y, z)$ from the velocity field. $f(x, y, z)$ is
given by $g(\x, \x_0, \delta \x) g(z, z_{\rm c}(\y), \delta z)$, where
$g(s, s_0, \delta s) = 4 / (1 + exp[(s - s_0)/\delta s])(1 + exp[(s_0
  - s)/\delta s])$ and $z_{\rm c}(\y) = z_0 + \Delta z \sum_{i=0}^{64}
\sin(2\pi i \y / 6 {\rm km} + \varphi)$, where $\varphi$ is a random
phase between $0$ and $2 \pi$. We use $\x_0 = 7.2$ km, $\delta \x =
187.5$ m, $z_0 = 1.51$ m, $\delta z = 10$ cm and $\Delta z = 10$
cm. This perturbation resemble a smooth bell-shape on the $\x$-$z$
plane, whose centre varies sinusoidally with $\y$.  The choice of $64$
for the last $\sin$ wavenumber is arbitrary, but the wavenumbers after
$\sim 10$ do not really contribute to the final result.

\section{The flame front instability}
\label{sec:res}

\begin{figure*}
  \twofigL{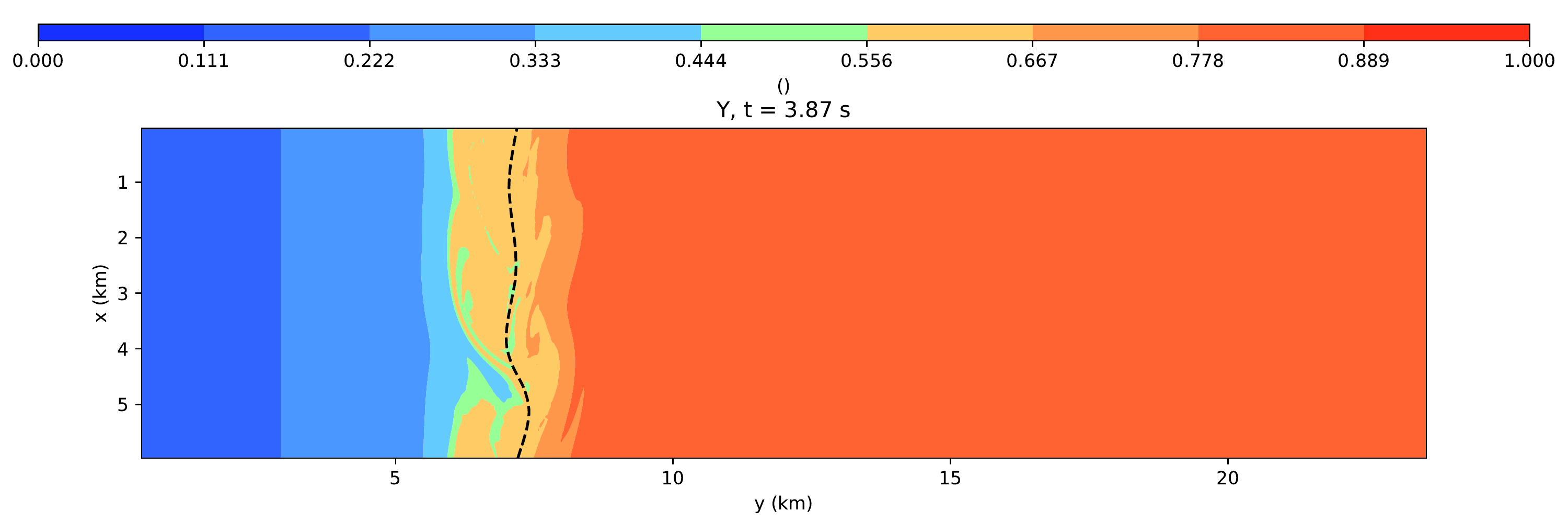}{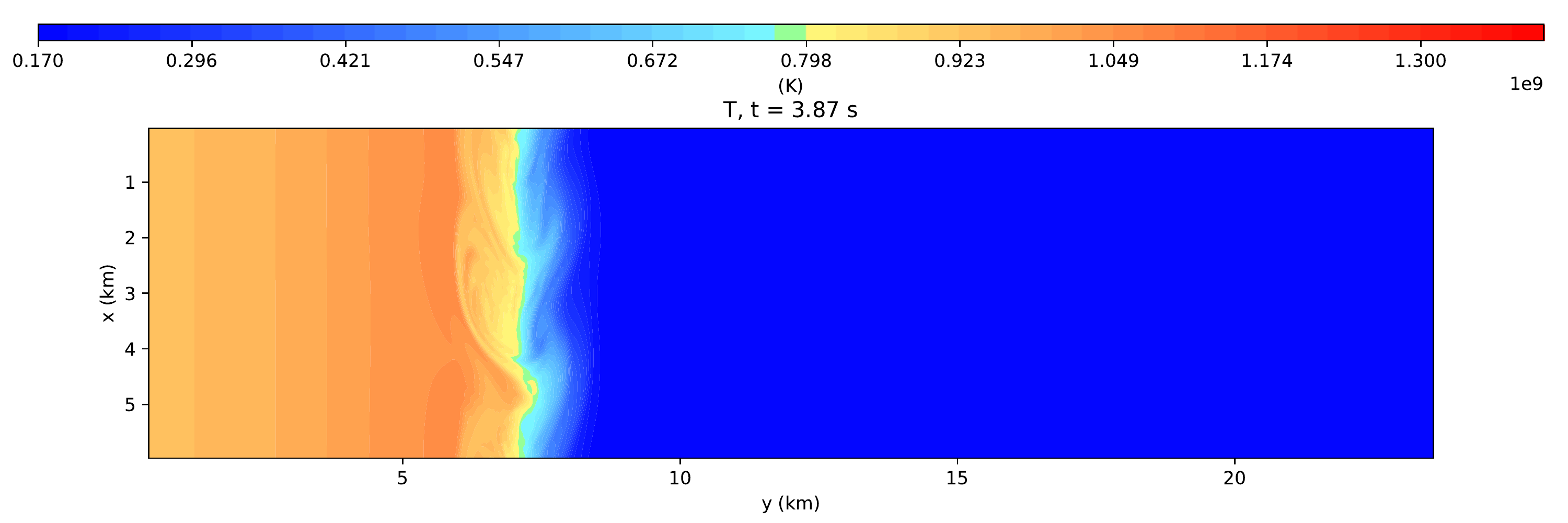}
  \caption{The beginning of the non linear regime of instability,
    horizontal slice at $H = 4$ m. \textbf{Top panel}: helium mass
    fraction $Y$ near the bottom of the burning layer. The fluid
    composition can be used to follow the fluid motion. The black
    dashed line indicates a pressure contour ($P < 2.2\tent{22}$ erg
    cm$^3$, compare also \figrefs{fig:instabY}, \ref{fig:instabvortA}
    and \ref{fig:instabvortB}). The highest pressure is on the leftmost
    part of the plot. \textbf{Bottom panel}: temperature profile.}
  \label{fig:instabini}
\end{figure*}

The left panel of \figref{fig:uyvertprofs} shows the initial
background conditions for the perturbed run. In particular, we plot
the azimuthal velocity profiles with filled contours and the density
stratification with dotted lines.  One can identify the thermal wind
profile and its vertical shear. Also note that where the density
contours (isopycnals) slope upwards is where the misalignment with
isobars peaks (and so does baroclinicity $\nabla P \times \nabla
\rho$): this is where we seed our perturbation (at the location of the
dashed vertical line: $\x = 7.2$ km).

The angle the isopycnals form with the horizontal is called the
\emph{wedge of instability} \citep{book-1987-Pedlo,book-2006-vallis},
because exchanging fluid elements within this angle brings lighter
elements among denser fluid and vice versa, and buoyancy then sustains
the perturbation leading to the instability. The overall centre of
mass of the fluid is lowered and gravitational potential energy is
converted into kinetic energy. This angle is very shallow: note that
the vertical length scale of \figref{fig:uyvertprofs} is just a few
meters while horizontally it is kilometres, so that the motion
following the instability is almost horizontal. Another aspect to
consider is the effect of the pure horizontal velocity shear, which
can also contribute to the growth of instabilities, even though
usually on smaller scales \citep[this is called the barotropic
  instability,][]{book-1987-Pedlo}.

This configuration and the development of the baroclinic instability
is a very active field of research in geophysical sciences. A direct
comparison with geophysical results is, however, difficult, because
there are no general models for the development of the instability
(especially for the non linear regime). Also, the systems studied in
the literature are either simplified few layer models or have
important differences from our simulation.  For example, we have an
important internal heat source (the burning), which reinforces the
conditions for instability and is not always present in the
geophysical models. When heat is included in those models, as in the
case of heat released by condensation, it usually takes places in the
top layers and not at the bottom like in our case.

One important point in our simulations regards the Coriolis force. We
model an $f$-plane \citep{book-1987-Pedlo}, which is a system with
constant Coriolis force, while in reality its intensity should change
with latitude. This effect is captured for example by the so-called
$\beta$-plane models, but in this exploratory work we neglect
it. Perhaps the most easily comparable common geophysical model is the
Eady model \citep{art-1949-Eady}, which consists of a uniformly
stratified ideal gas, confined by two rigid lids on an $f$-plane, with
uniform vertical shear. In this model, the growth rate of the most
unstable mode is given by \citep[Eq. 9.87 of][]{book-2006-vallis}:

\begin{equation}
  \sigma = U \frac{k}{\mu}\left[
    (\coth \frac{\mu}{2} - \frac{\mu}{2})
    (\frac{\mu}{2} - \tanh \frac{\mu}{2})
    \right]^{1/2}
  \label{equ:eady}
\end{equation}
where $\mu = R_{\rm Ro} (k^2 + \pi^2 / L^2)^{1/2}$. $U$ is the
velocity of the top layer and $L$ the width of the channel where the
instability takes place. While simplified, this model captures most of
the features of the instability \citep{book-2006-vallis} and we will
briefly compare our results to it. \appint{}

\subsection{The early stages of the instability}

We have measured the growth rate of the instability during the linear
phase. This is shown on the right panel of \figref{fig:uyvertprofs},
where we over-plot the prediction of \eqr{equ:eady} with parameters,
taken form the unperturbed simulation, $R_{\rm Ro} = 5\tent{4}$ cm, $U
= 10^8$ cm s$^{-1}$, $L = 3 R_{\rm Ro}$. This choice is approximated,
and somewhat ad hoc, but the qualitative picture agrees.  The fastest
growing mode has a wavelength of $\sim 2$ km, which corresponds to
approximately $\lesssim 4$ Rossby radii and the shortest wavelengths
are stable\footnote{Had we included $\beta$ effects, there would have
  been also a cut-off at longer wavelengths.}. This compares
qualitatively well with what is predicted by the Eady model if one
considers that the analogy is not complete since our conditions are
not exactly the same (for instance, the top layer is not a rigid
lid). Thus, we can consider this a robust result and expect that the
fronts on the surface of neutron stars will corrugate when they exceed
a few Rossby radii in length.

\begin{figure*}
  \twofigL{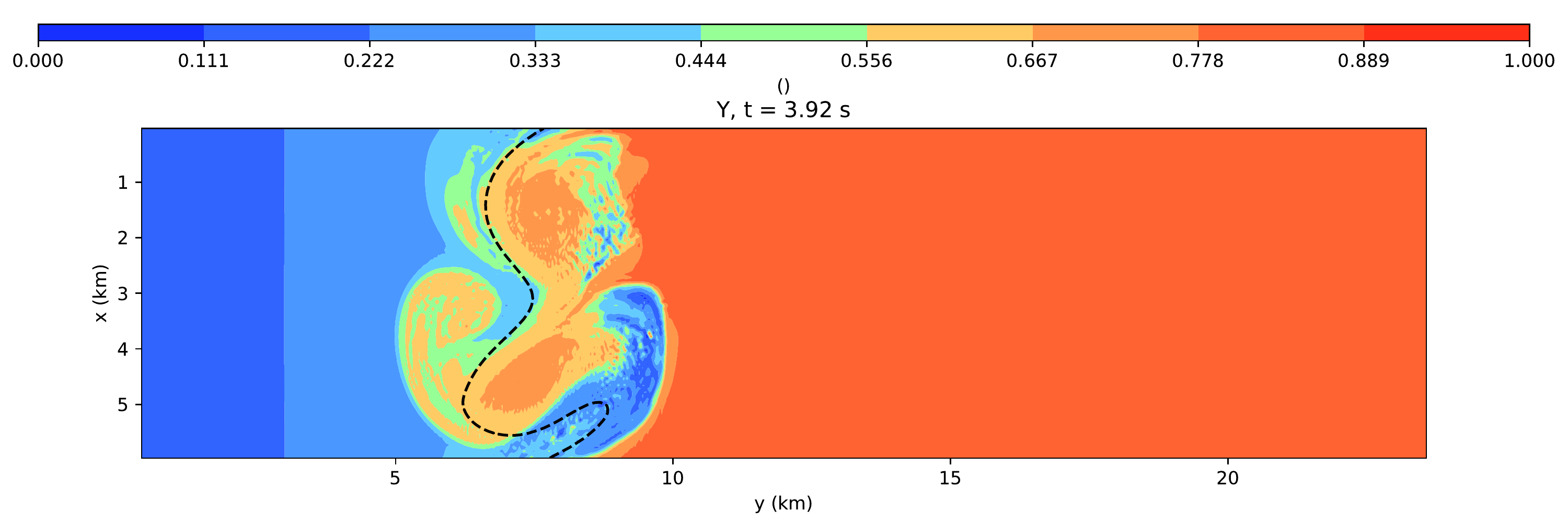}{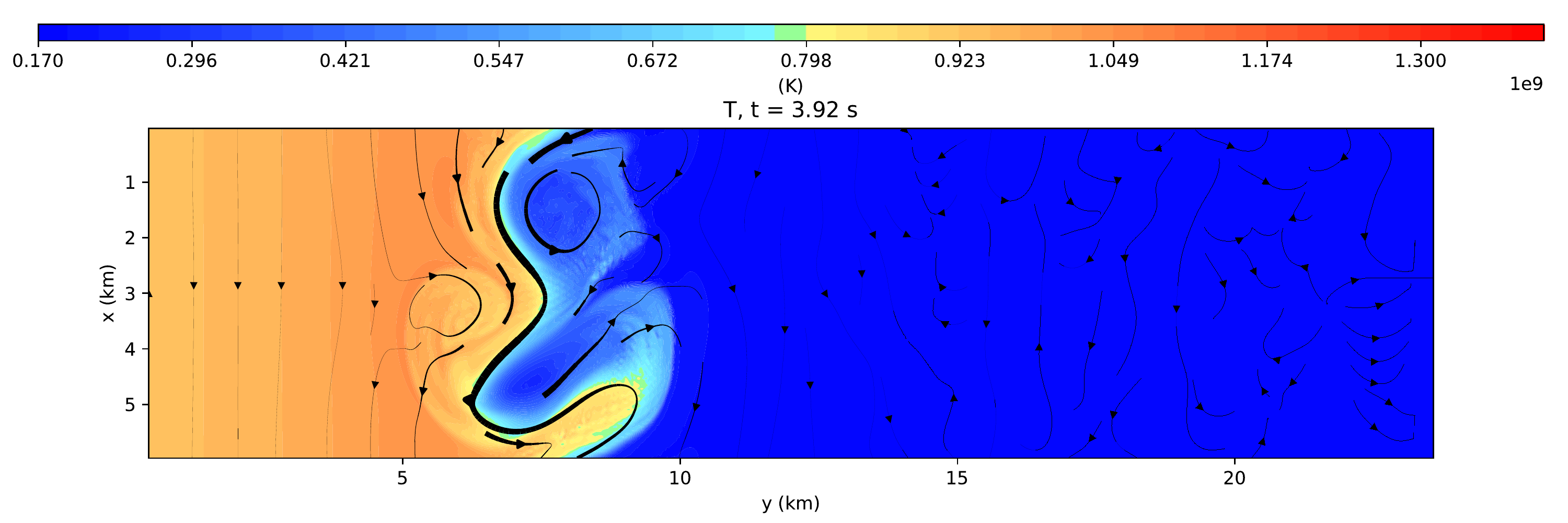}
  \caption{Same as \figref{fig:instabini}, at later time. For clarity,
    from now on we also add streamlines on top of the temperature
    profile. The thickness is proportional to the local
    speed. Different figures will have different scale. The fluid
    motion leads to both cyclonic and anti-cyclonic structures. The
    transport of hot fluid northwards and combustible fluid southwards
    is now fully evident.}
  \label{fig:instabY}
\end{figure*}

\figrefs{fig:instabgrow} and \ref{fig:instabini} show the instability
at the end of the linear phase--beginning of the non linear one. In
the left panel of \figref{fig:instabgrow} we plot vertical, azimuthal
slices, at the location of the dashed line of
\figref{fig:uyvertprofs}, for the temperature and the meridional and
vertical velocities. For each field we subtract the initial profile to
highlight the effect of the instability. The right panel shows
streamlines on horizontal slices at various heights. The red lines
follow the unperturbed flow, while the black lines show the
residuals. To guide the eye, the thickness of the lines is
proportional to the local speed of the streamline, but for visibility
the scale is different for each group of lines.

The most apparent feature is the typical \emph{leaning} of the
velocity contours against the background flow. Also, note that the
north-upwards perturbation flows correspond to the higher temperatures
and the south-downwards ones to the colder fluid: there is a net flux
of heat towards the pole.  Analogously, the flow is carrying
combustible fuel south and more ashes-rich fuel north. This
configuration is typical for the instability and compatible with what
we saw in the 2.5D simulations of \citet{art-2013-cavecchi-etal,
  art-2015-cavecchi-etal}. Also in the 3D run we identify this trend
as a contribution to the speed up of the flame.

It can be seen that at these stages the dominating mode has wavenumber
$5$, but asymmetries in the peaks of the contours show that the mode
with wavenumber $1$ is quickly catching up while the instability is
saturating. This begins to be evident already in
\figref{fig:instabini}, which is a snapshot taken only $0.01$ s later,
and become manifest in \figref{fig:instabY}, which shows the evolution
in the fully non linear stage. The development is reminiscent of the
mushroom shape of the Rayleigh-Taylor instability, but the motion is
almost along a horizontal direction. The fluid slides along the
isopycnal surfaces and breaks into cyclonic and anticyclonic systems:
see the systems respectively to the right and to the left of the
dotted line corresponding to an isobar on the top panel of
\figref{fig:instabY} and compare to the streamlines drawn on the bottom
panel\footnote{Cyclonic systems are centred on low pressure and rotate
  clockwise in the northern hemisphere (as in our simulation
  domain). Anticyclonic systems are centred on high pressure and
  rotate counterclockwise in the north.}.

\begin{figure*}
  \twofig{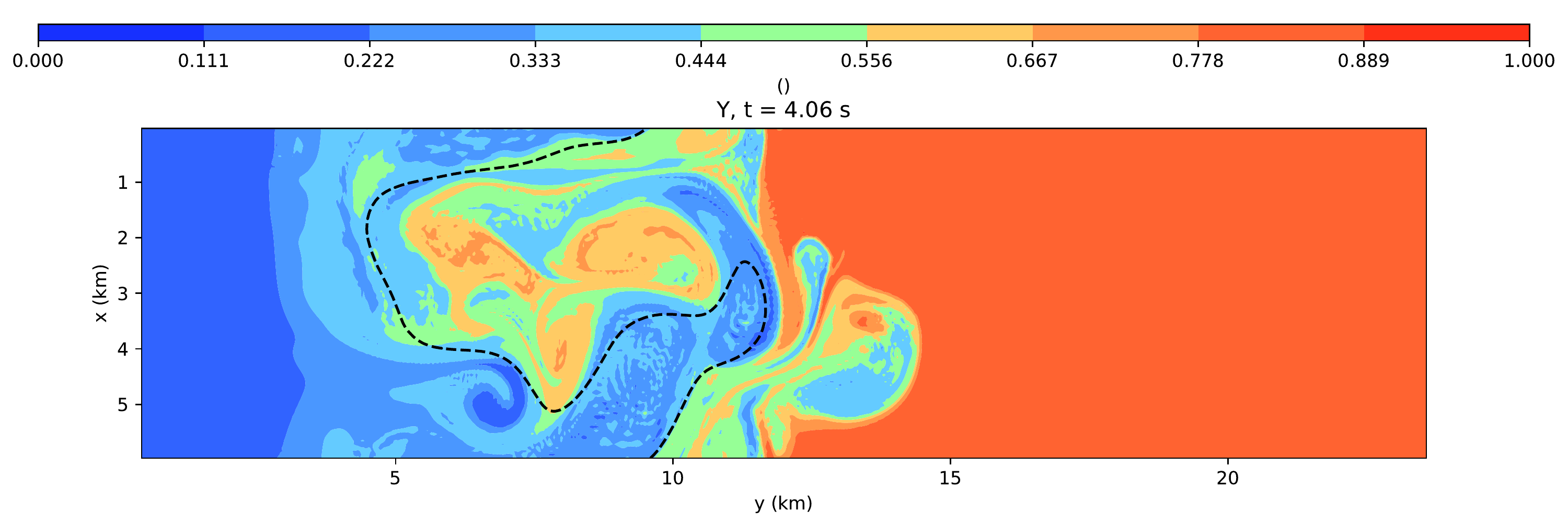}{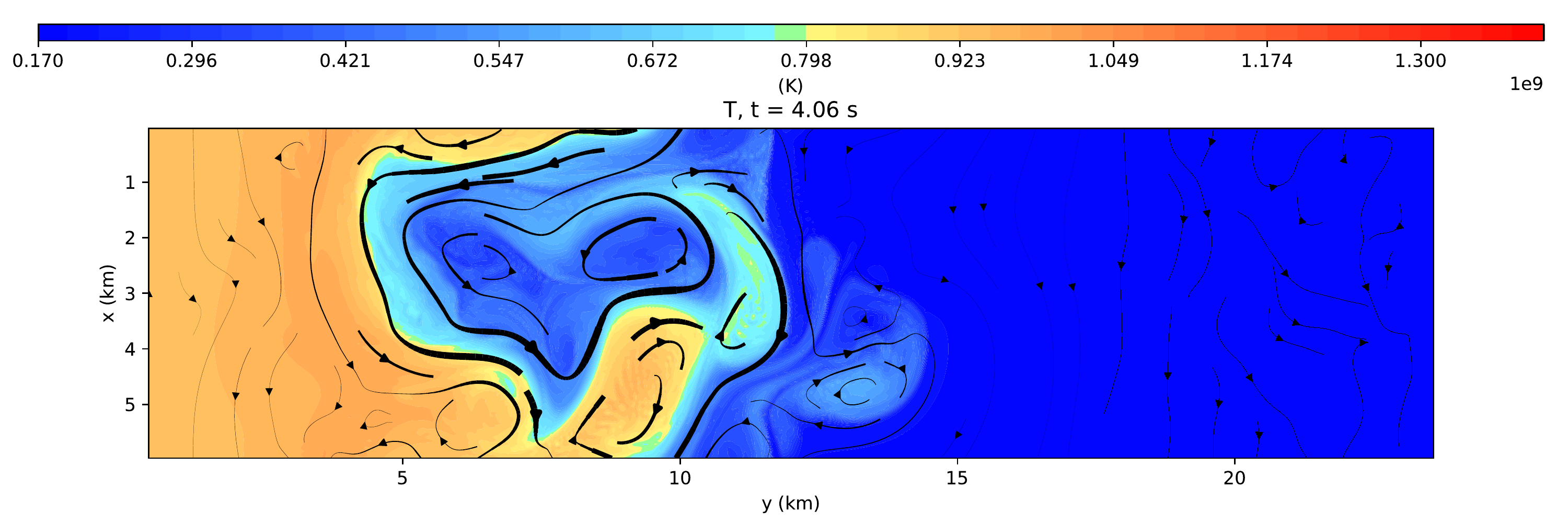}
  \twofig{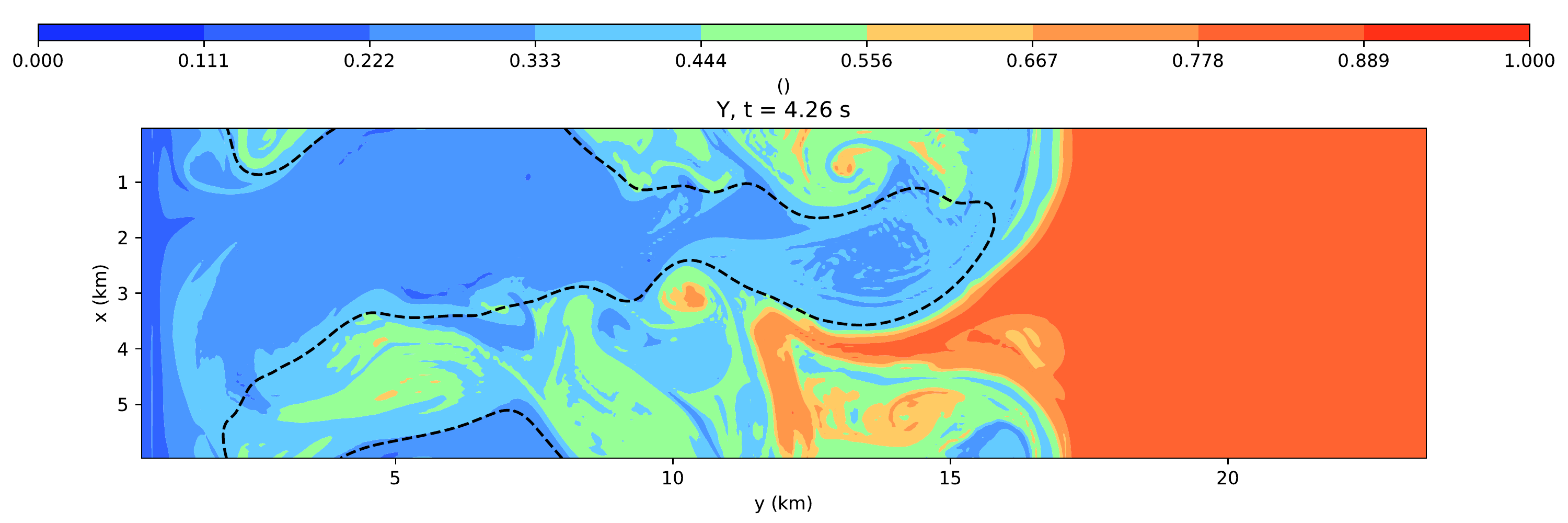}{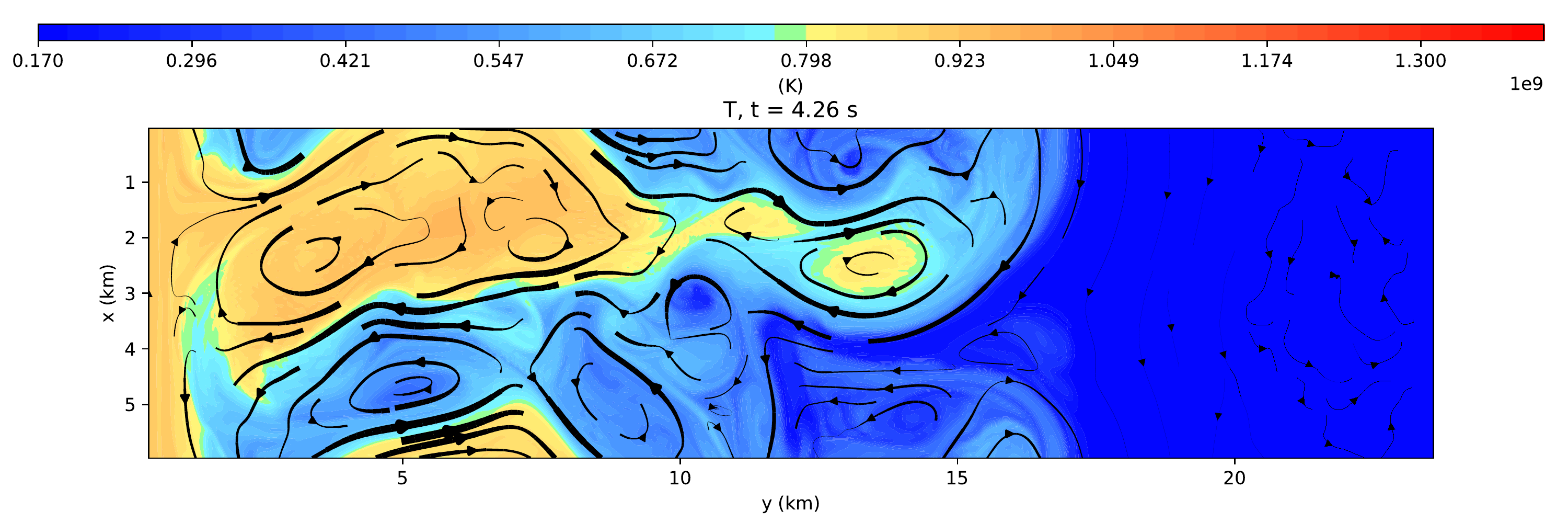}
  \twofig{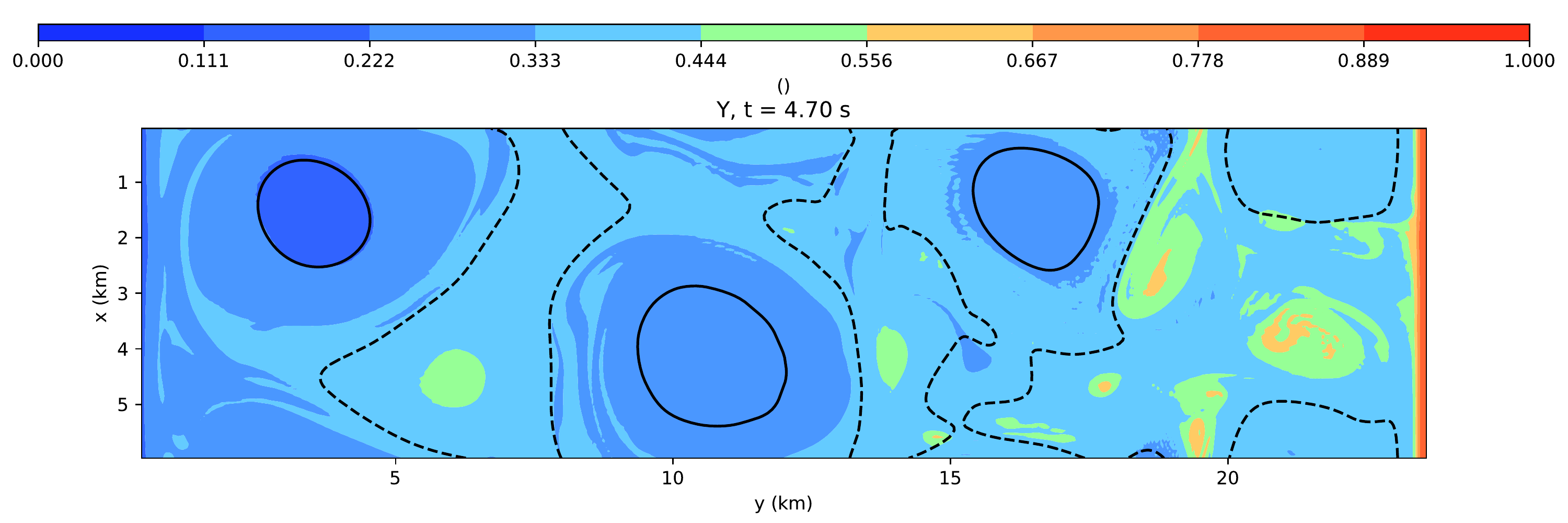}{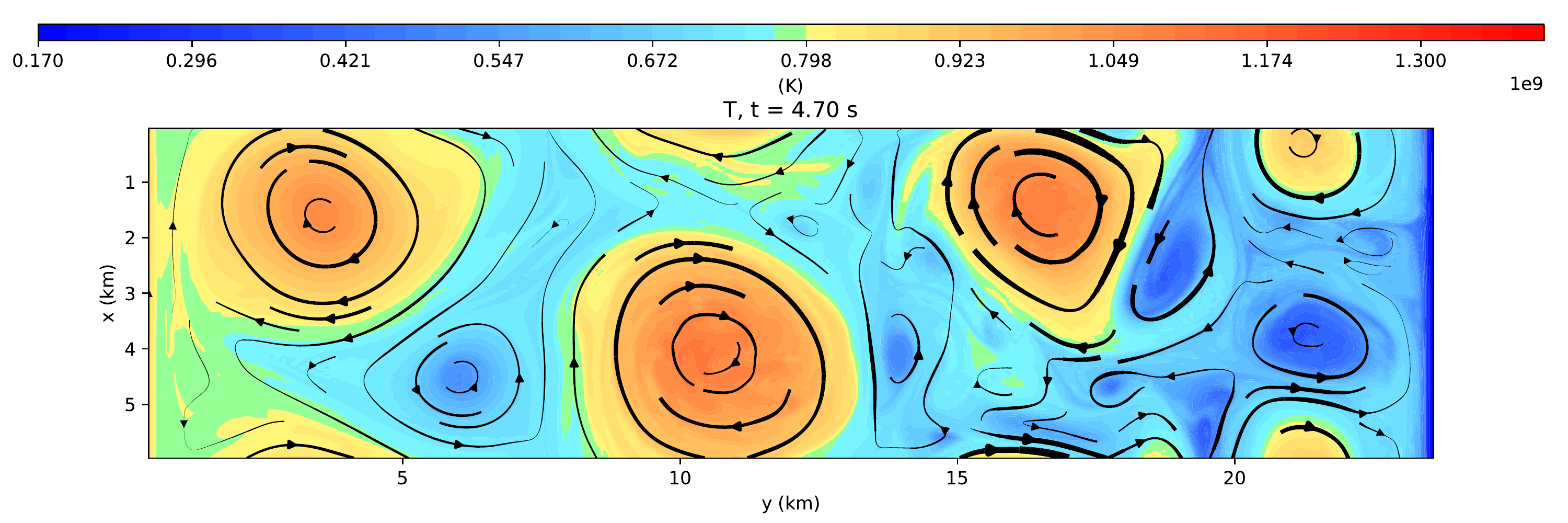}
  \caption{Vortices in the last stages of flame propagation. As in
    \figref{fig:instabini} the panels are slices at $H = 4$
    m. \textbf{Left panels}: helium mass fraction. Black lines are
    isobars (dotted if $P < 2.2\tent{22}$). \textbf{Right panels}:
    temperature. The flame reaches the end of the domain $\sim 10$
    times faster than in the unperturbed case.}
  \label{fig:instabvortA}
\end{figure*}

\subsection{Later stages and the development of giant vortices}

\begin{figure}
  \begin{center}
    \includegraphics[width=0.47\textwidth]{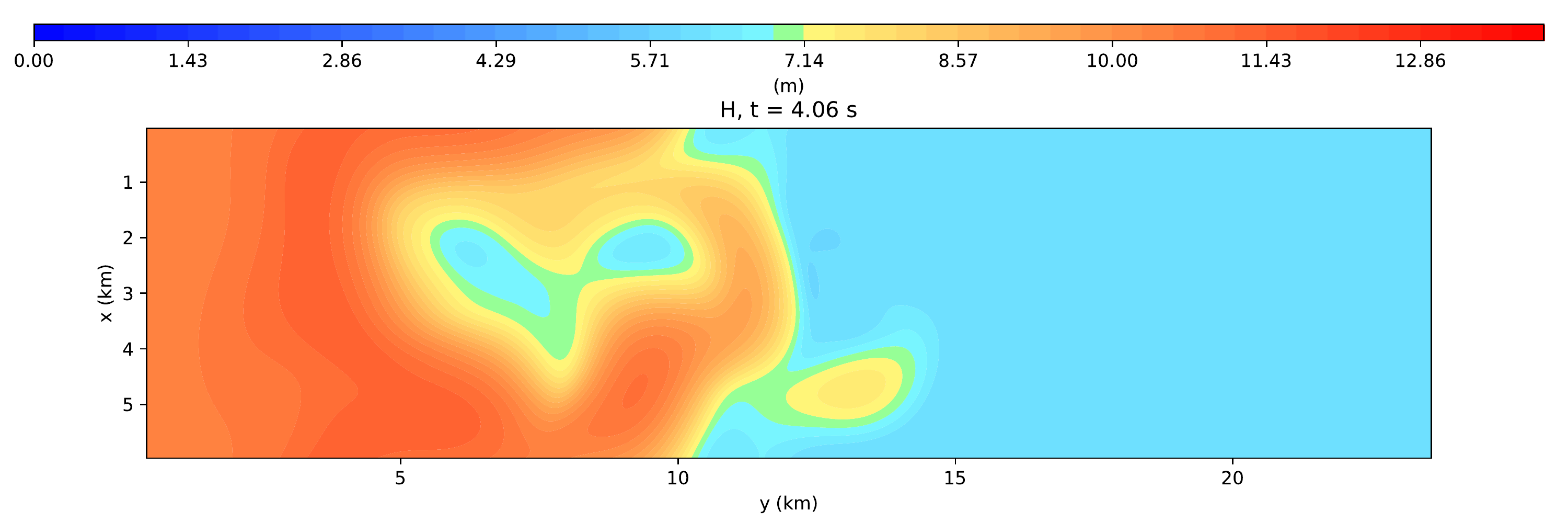}\\
    \includegraphics[width=0.47\textwidth]{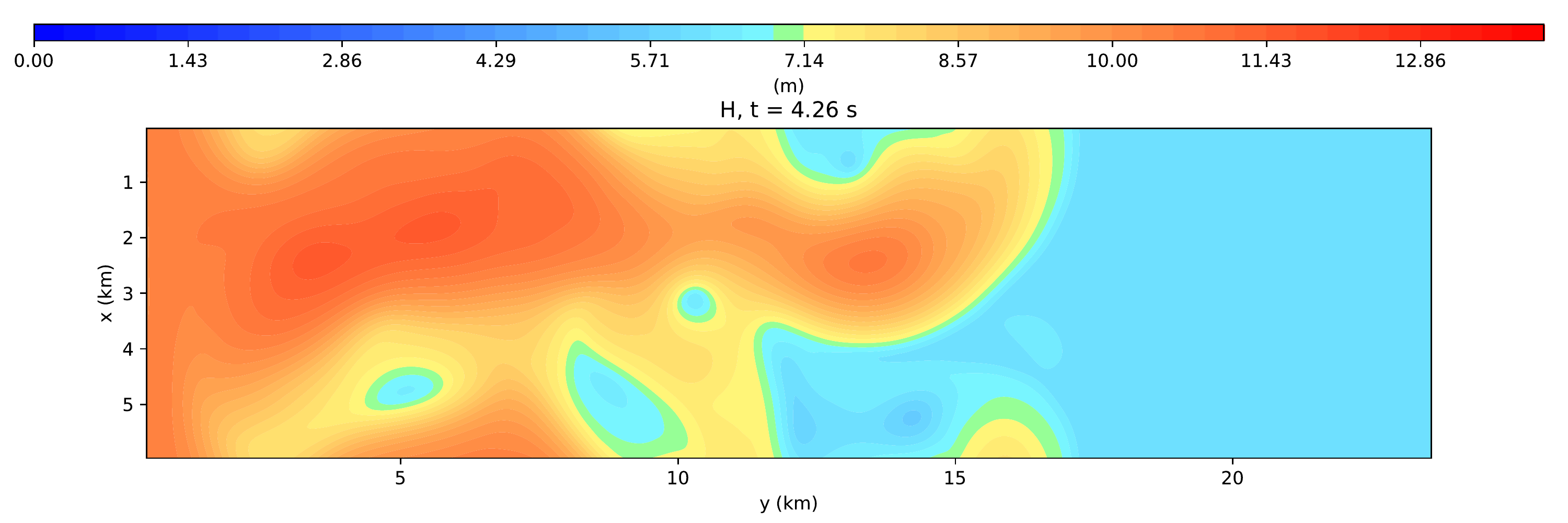}\\
    \includegraphics[width=0.47\textwidth]{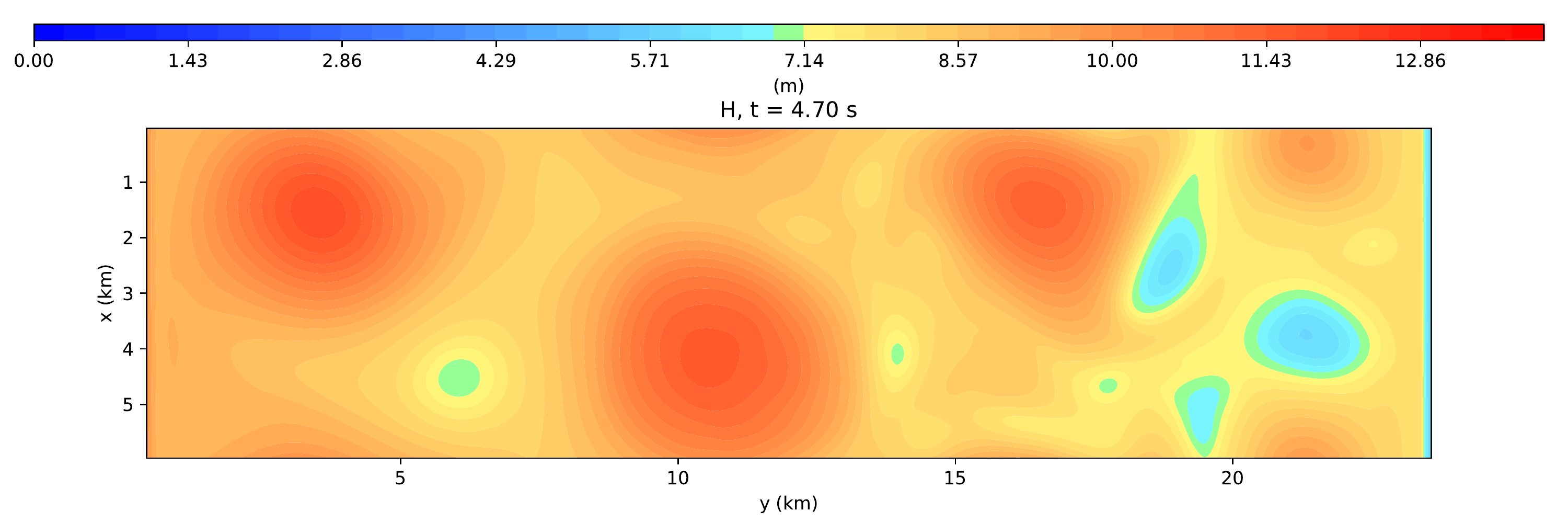}
  \end{center}
  \caption{Vortices in the last stages of flame propagation.  Total
    height of the simulation at the same times as in \figref{fig:instabvortA}}
  \label{fig:instabvortB}
\end{figure}

The net effect of the flows described above is to trigger circulation
on top of the original azimuthal flow and to cause ignition at various
locations northwards. Eventually, various systems of locally
higher-temperature higher-pressure and lower-temperature
lower-pressure form at alternate $\x$ coordinates (corresponding to
higher latitudes) creating a complex combination of anticyclones and
cyclones (see \figrefs{fig:instabvortA} and \ref{fig:instabvortB}).

Bigger systems drift and interact with smaller vortices absorbing or
disrupting them. One large anticyclone, approximately $4$ km across
($\sim 8 R_{\rm Ro}$), is fully formed near the left at $\x \sim 4$
km, $t = 4.25$ s ($\sim 0.5$ s after the perturbation was
triggered). A second anticyclone, which originated from a small system
detached roughly at $t = 4.07$ s from the first one, is clearly
recognisable at $t = 4.48$ and $\x \sim 11$ km. This one in turn
launched another small system at $t = 4.38$ s which eventually becomes
a third large eddy identifiable at $t = 4.69$ s and $\x \sim 16$
km. The ``launching'' happens when the the major systems enter in
contact with a smaller cyclonic system. Due to the mixing effect of
this interaction, the flame reaches the north boundary very
quickly. The fact that we observe three such systems is probably due to
the fact that only three could fit within the domain.  Different stars
spinning at different frequencies will most probably show a different
number of such anticyclones.

By late times (bottom rows of \figrefs{fig:instabvortA} and
\ref{fig:instabvortB}) the three large anticyclonic systems are stable
and most of the burning is localised at their core. A few cold
cyclonic systems are still present between them and at the northmost
latitudes, but the flame rapidly fills the gaps between the burning
systems. Their centres slightly drift, ``guided'' by the background
flow and the interaction with other sub-systems, mostly towards the
East. A more detailed discussion of the late evolution of the
vortices, their drifts and the effects on the observed lightcurves
will be presented in future work.

\section{Discussion}
\label{sec:discu}

The first conclusion to be drawn is that, despite the quick and highly
non-linear evolution of the instability, the flame is able to survive
until the end of the domain, which, as we said, is comparable to the
equator-pole distance of a somewhat large star of $15$ km of
radius. We can then compare the perturbed flame results to the 2.5D
simulation ones.

\begin{figure*}
  \includegraphics[width=0.45\textwidth]{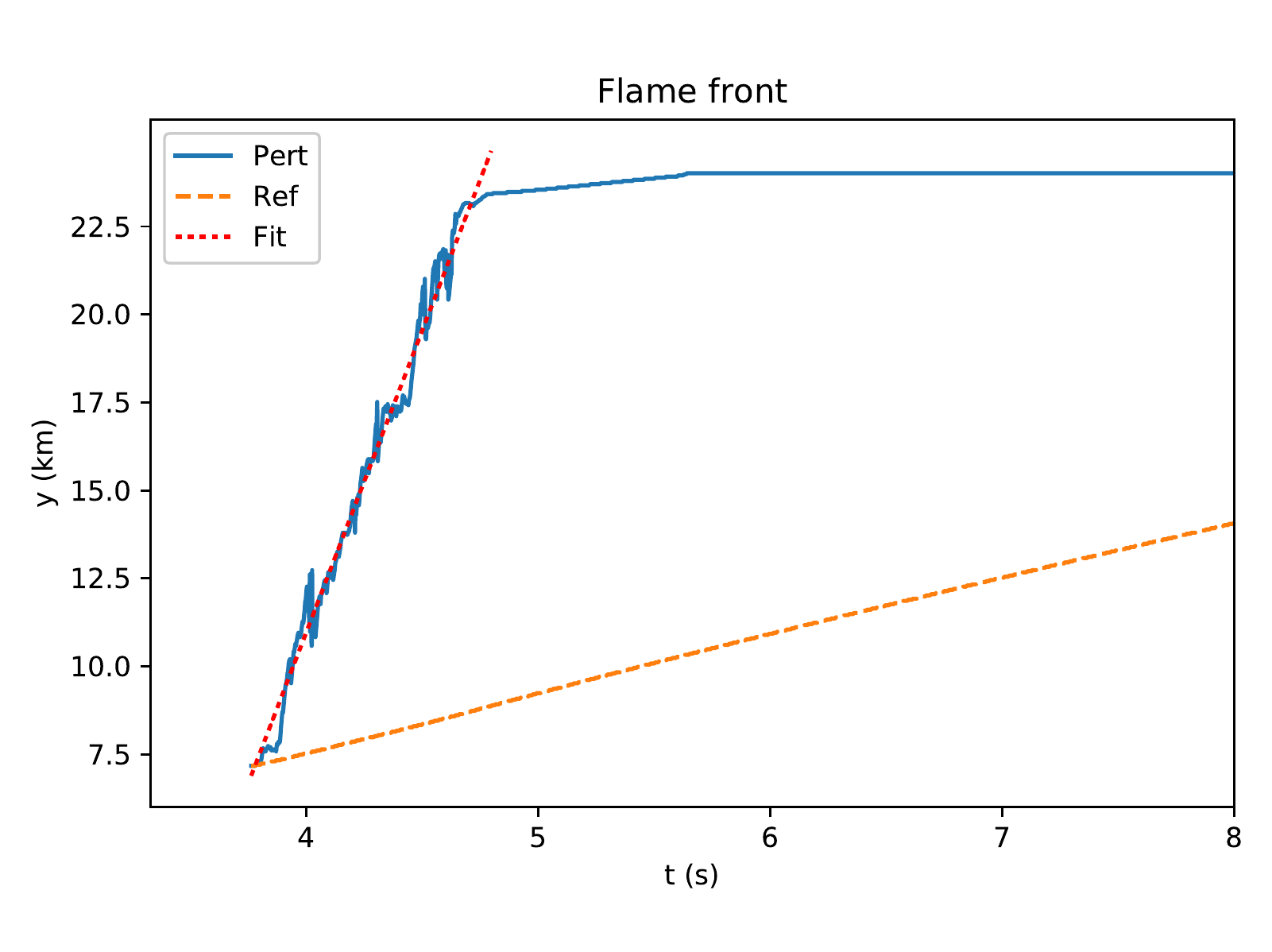}\hspace{\stretch{1}}
  \includegraphics[width=0.45\textwidth]{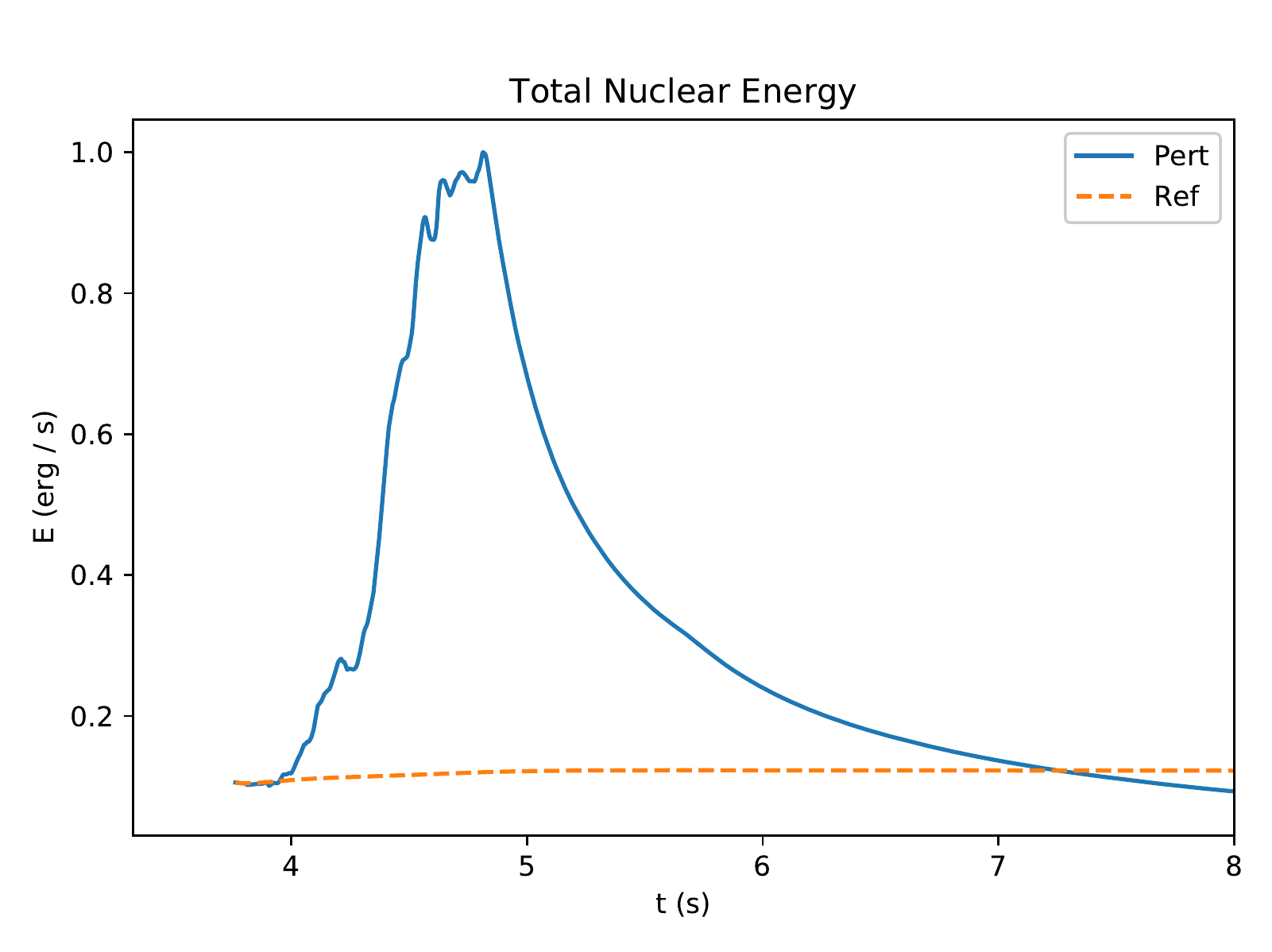}
  \caption{Comparison of the global features of the perturbed flame
    versus the unperturbed one. \textbf{Left panel}: Flame
    velocity. \textbf{Right panel}: Volume integrated energy
    output (these curves are normalised to the maximum of the output
    from the perturbed run).}
  \label{fig:comparison}
\end{figure*}

The first thing to compare is the global velocity of the flame. To
measure the speed of the flame we have written a front-tracking tool
which identifies the position of surfaces with a given value of the
temperature and then returns the most advanced position. This is
needed because the presence of the various anticyclones makes other
methods very noisy.  We set the tracking temperature to $7\tent{8}$ K,
since this is coincident with the beginning of significant helium
burning. The position of the flame front in the perturbed and
unperturbed case is shown in the left panel of
\figref{fig:comparison}. The comparison is striking: the unperturbed
velocity is $1.65\tent{5}$ cm s$^{-1}$, while the perturbed one is
$1.72\tent{6}$ cm s$^{-1}$. \emph{This means a speed up of $\sim 10$
  times}. This is a very interesting result, because it can explain
fast rise times for the bursts also in stars with a weak magnetic
field. The flame speed is comparable to the speed obtained by
\citet{art-2016-cavecchi-etal} in the case of a star with the same
spin and a magnetic field of $10^8$ G, $1.61\tent{6}$ cm s$^{-1}$,
which was the highest achieved for a star with this spin.  Since in
the simulations of \citet{art-2016-cavecchi-etal} the interaction
between magnetic field and the flow of the fluid lead to significant
changes, exploring the interaction of the instability and the vortices
with the magnetic field seems a natural next step. Another important
point we will explore is the dependence of the speed up on the spin of
the star alone. We can expect the factor to depend on $\Omega$ since
the mechanism leading to the speed up is the baroclinic instability
and the presence of the vortices. The scale for both is set by the
Rossby radius, which scales with $1 / \Omega$, so that we can expect a
similar scaling for the speed up. If we compare the total energy
produced within the simulation domain (\figref{fig:comparison}, right)
we can see that the increase of speed is reflected in the total energy
released. This is to be expected and easily explained by the early
ignition of the most distant fluid thanks to the \emph{almost
  horizontal} convection and mixing triggered by the instability and
the vortices.

\begin{figure*}
  \trifig{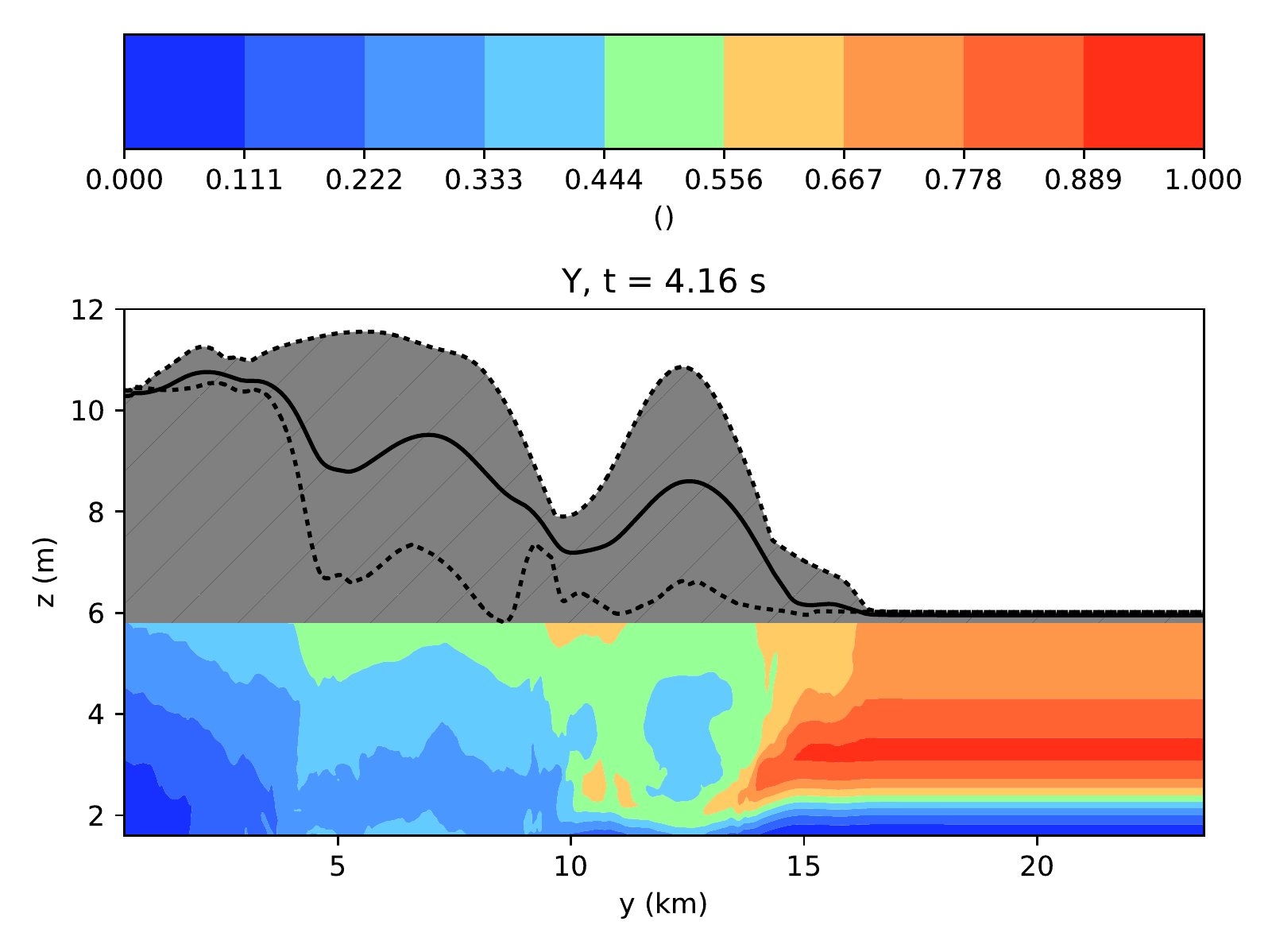}{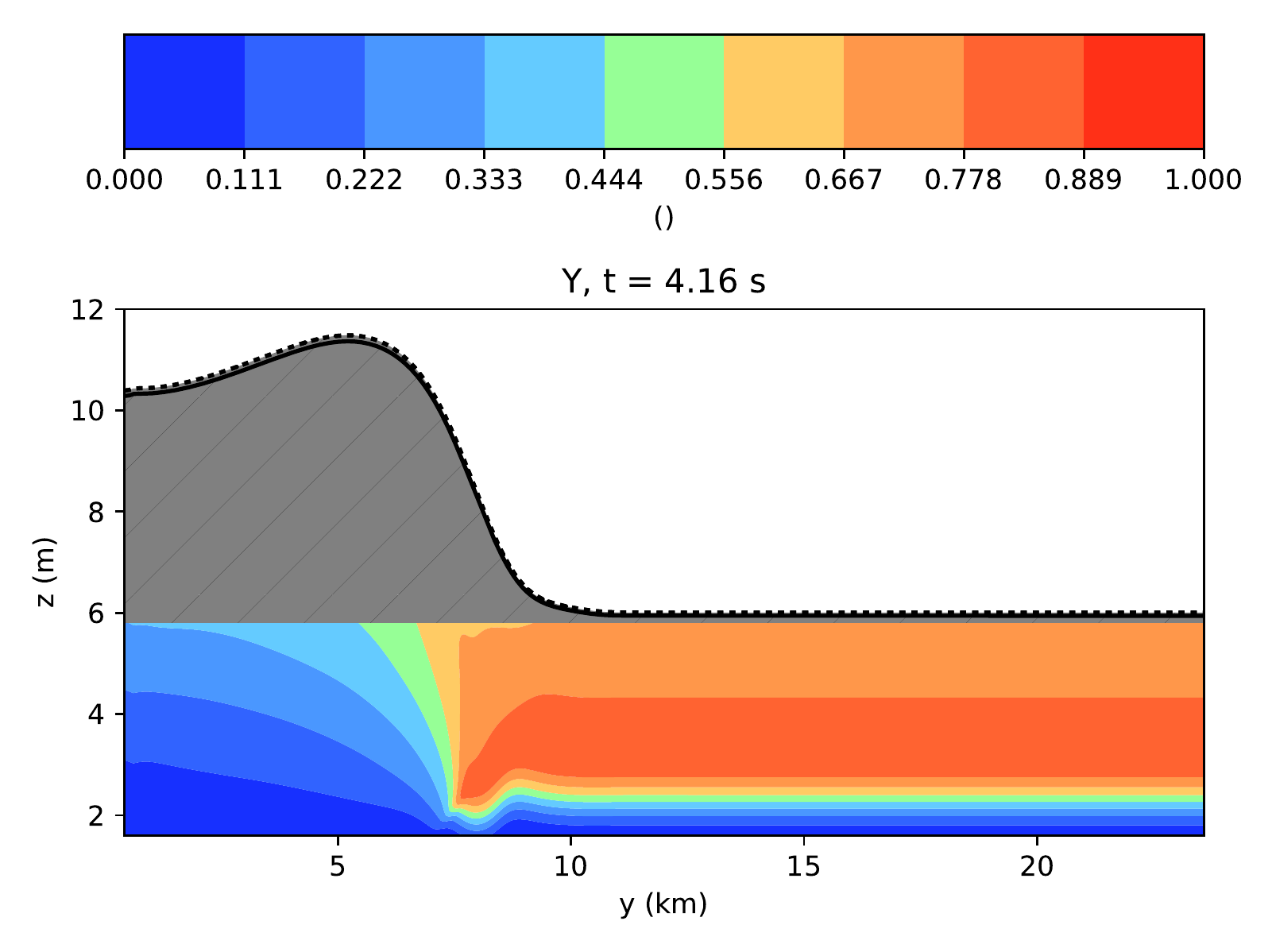}{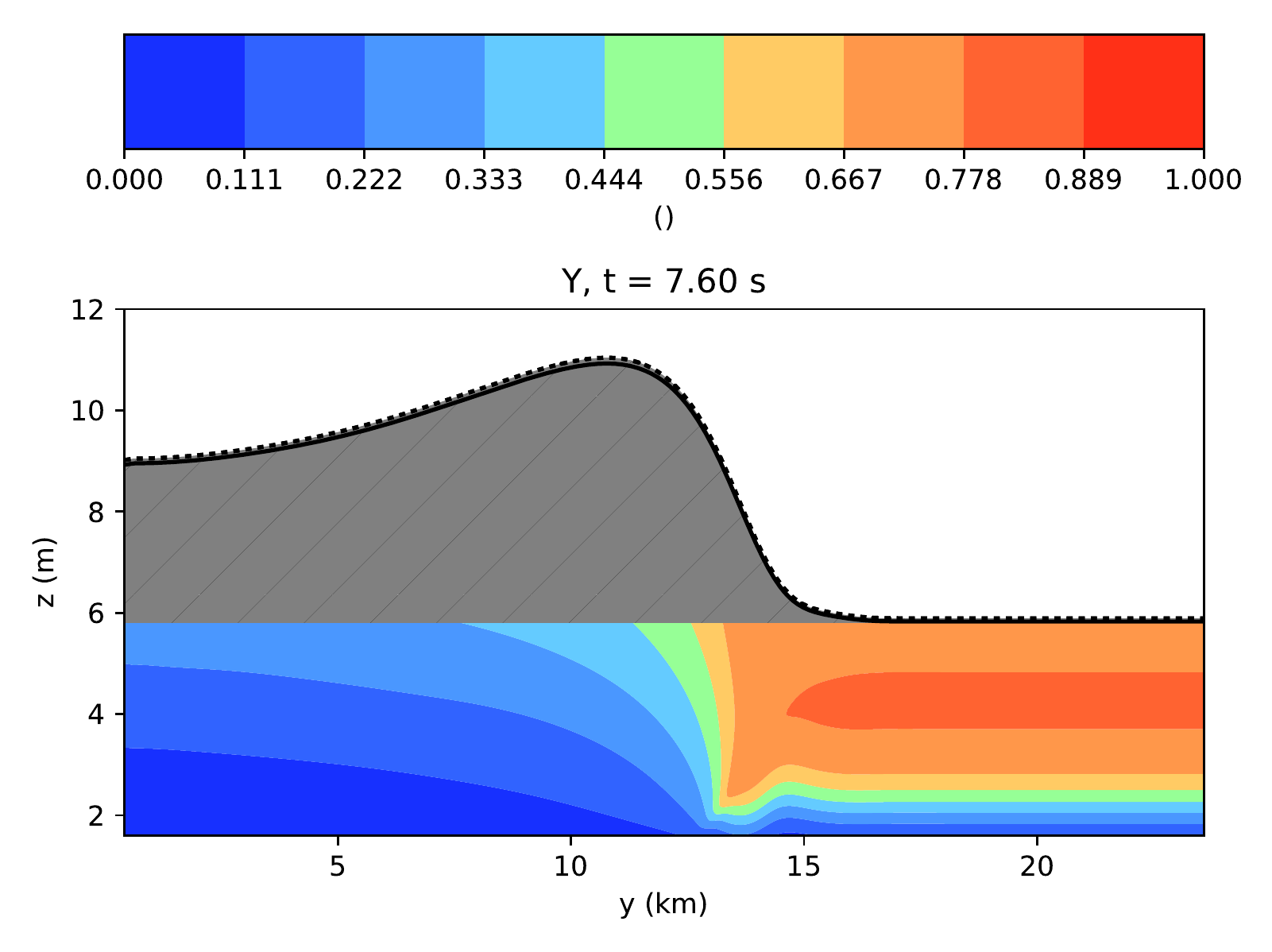}
  \trifig{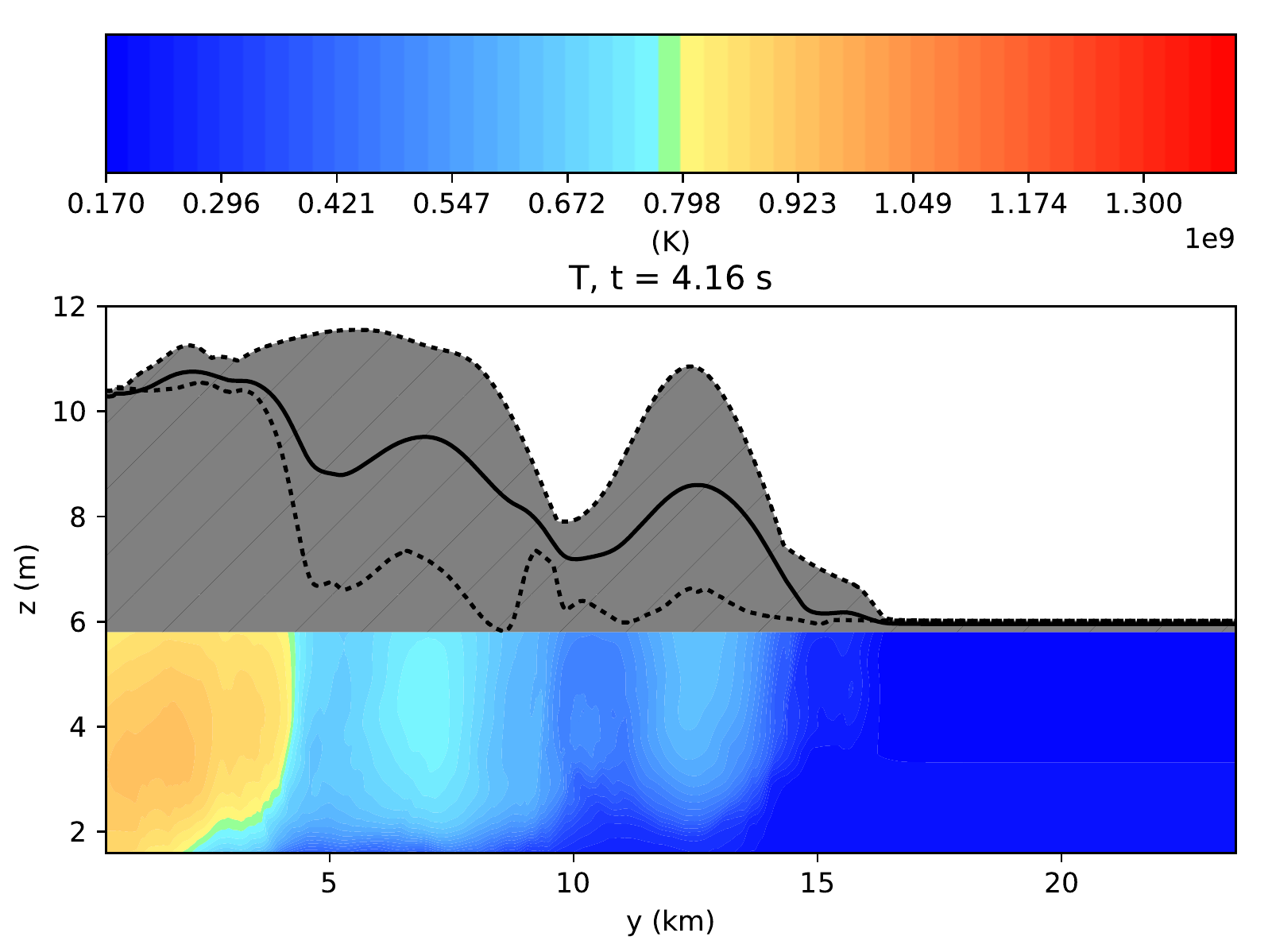}{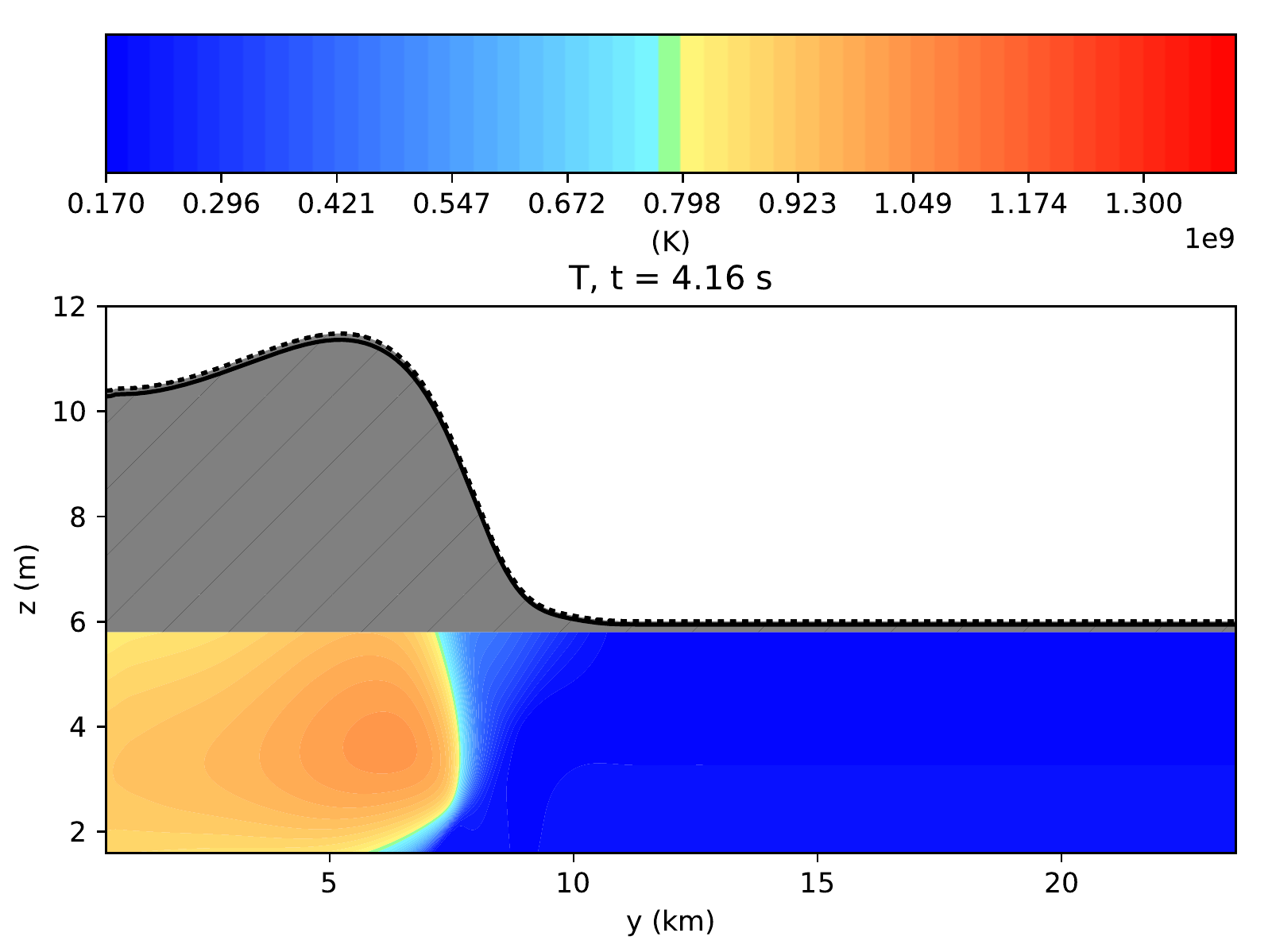}{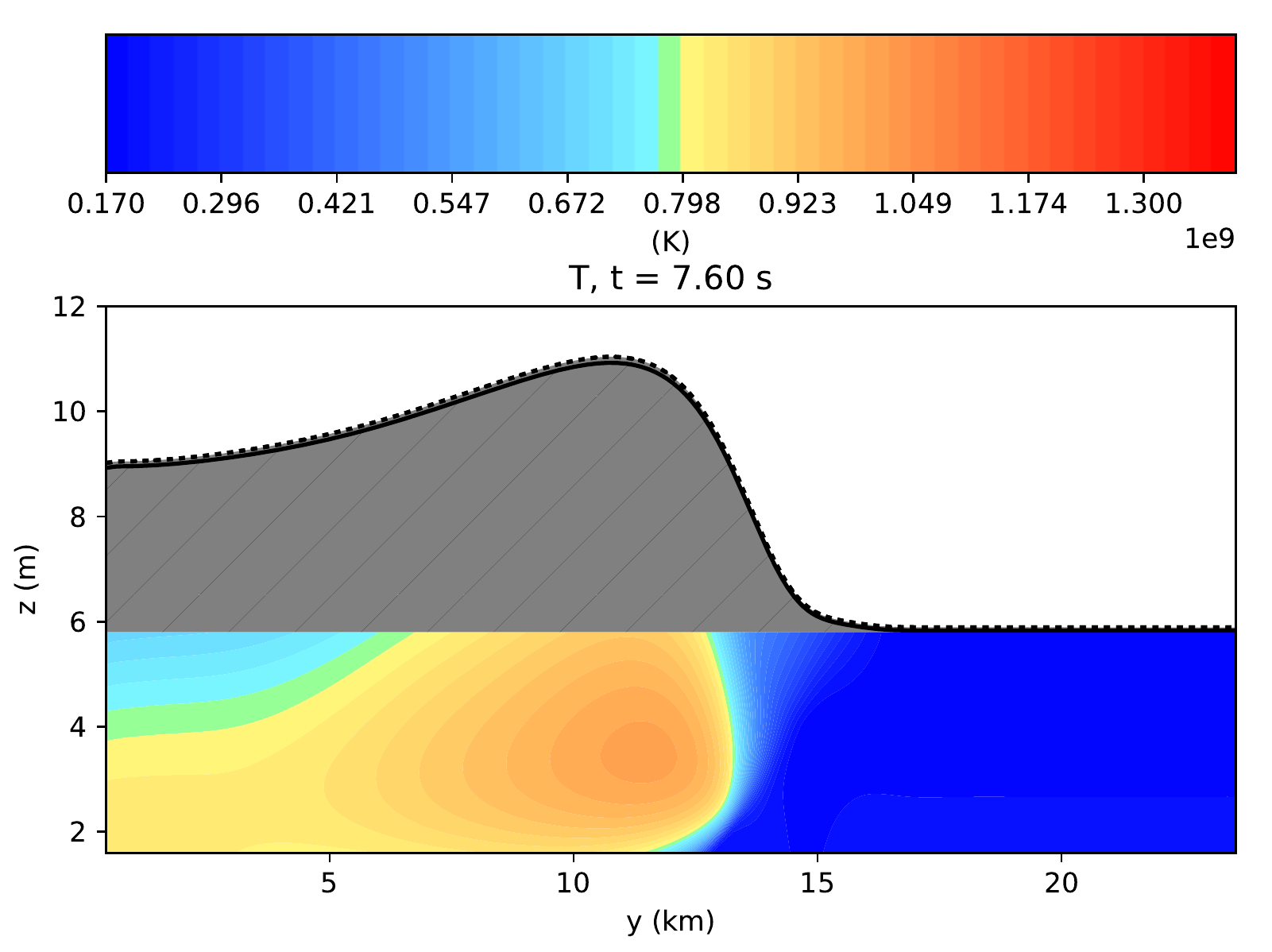}
  \trifig{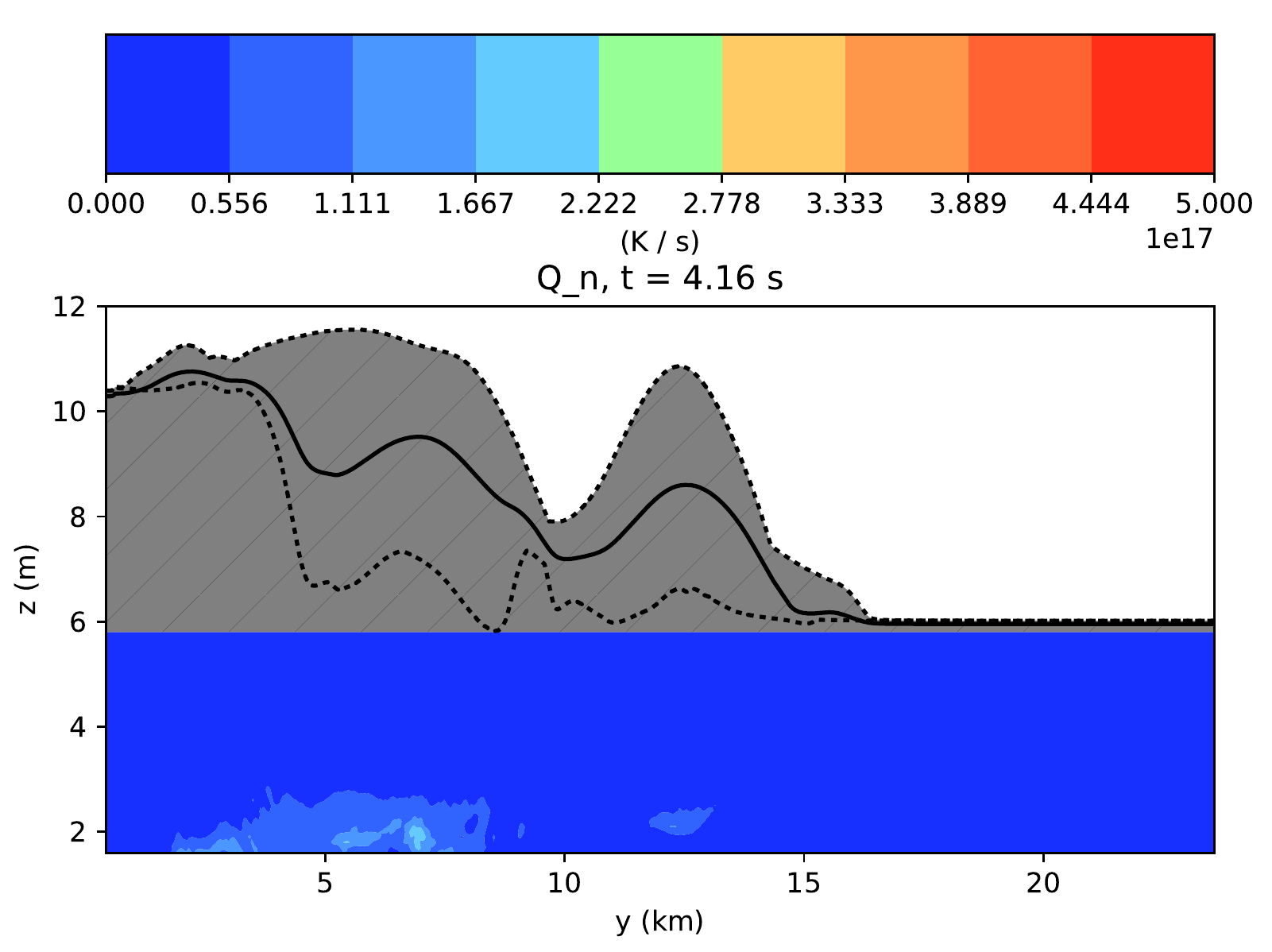}{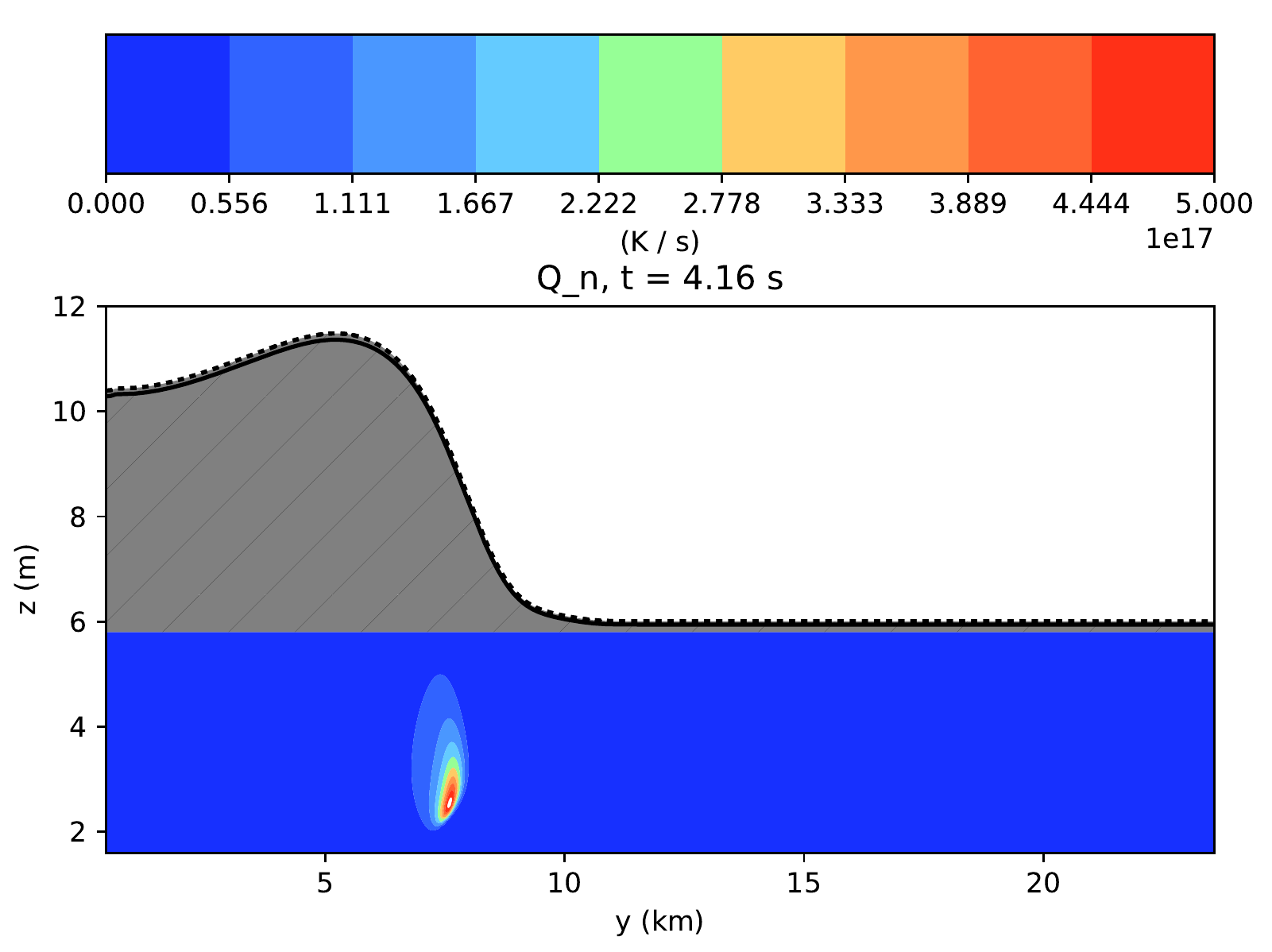}{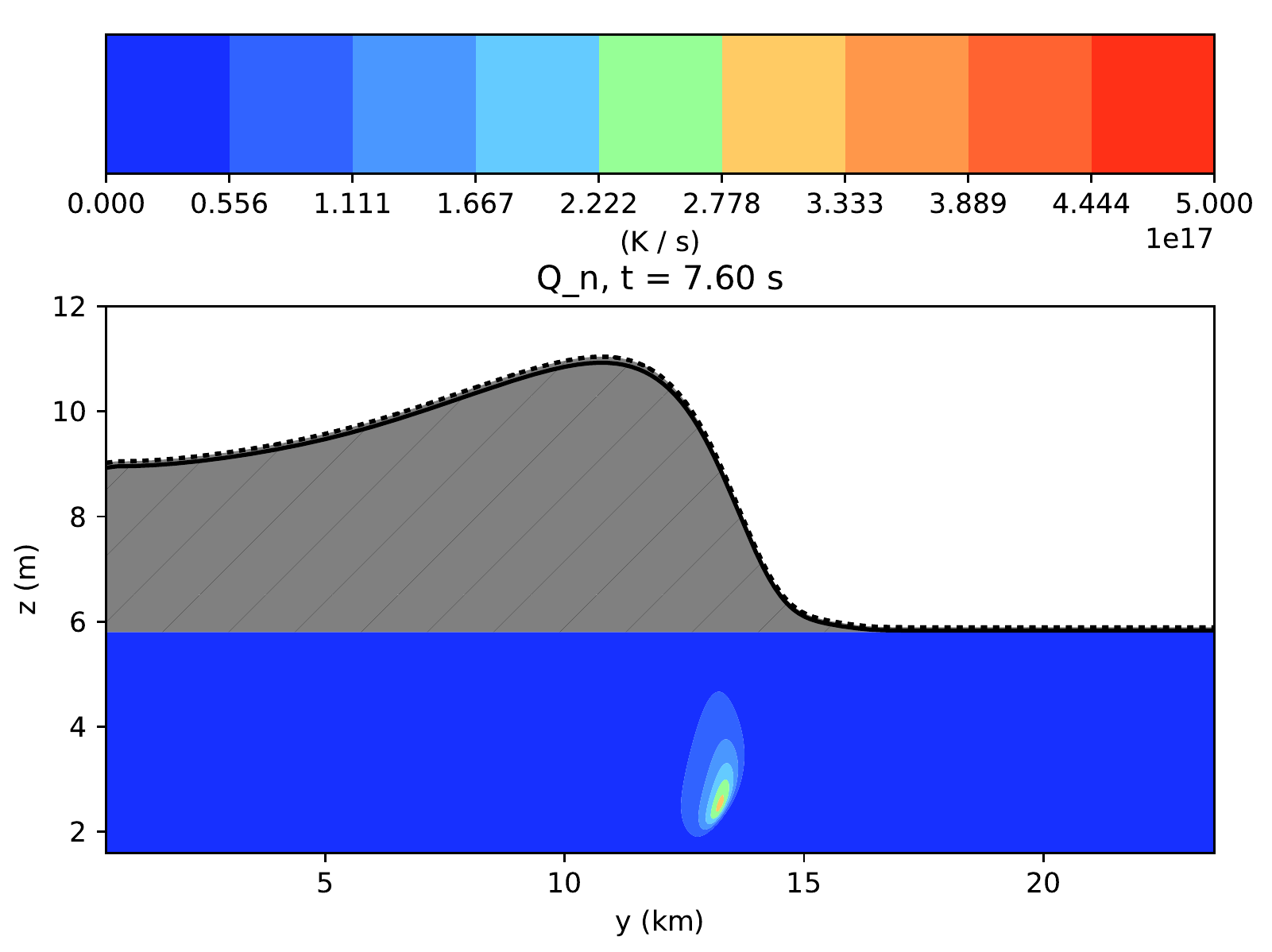}
  \caption{Comparison of 2.5D simulation with the azimuthal average of
    the perturbed simulation. Only a fraction of the domain is shown,
    both horizontally and vertically. The shaded region is not shown
    because it is impossible to calculate meaningful averages there in
    the 3D case. The top lines indicate the averaged height of the
    fluid (solid) and the minimum and maximum heights in the $\y$
    direction (dashed) for the 3D case. \textbf{Left panels}:
    azimuthal averages for the helium mass fraction, temperature and
    nuclear reaction yield for the perturbed solution. \textbf{Middle
      panels}: same fields for the 2.5D simulation at the same
    time. \textbf{Right panels}: same fields for the 2.5D simulation
    when the flame is at a position comparable to the one of the
    perturbed run.}
  \label{fig:compa2D}
\end{figure*}

Secondly, we can extract azimuthal averages from the 3D run, so to
have 2D profiles to compare to the 2.5D simulation. These averages are
shown in \figref{fig:compa2D}, where we plot the helium mass fraction
$Y$, the temperature $T$ and reaction yield $Q_{\rm n}$.  We compare
the perturbed flame simulation (first column) to two situations in the
unperturbed case: first to the 2.5D run at the same time (middle
column) and then to the situation at a (much later) time when the
flame has reached a position comparable to the one of the 3D
simulation (last column). Note that we cannot calculate averages at
every height, since we do not have data points above the top of the
simulation and in 3D the top layer is highly corrugated by the
vortices (as seen in \figref{fig:instabvortB}). For this reason, we
calculate the averages up to $5.8$ m, where data are available for all
the $\y$ domain. However, we over-plot the profile of the average
height as a black solid line and the profiles of the minimum and
maximum heights as dashed black lines to guide the eye.

First of all, some of the differences between the first and last
column are due to the fact that the time for the flame to reach $\sim
12$ km is longer, so that near the origin the fluid had time to cool,
while the fluid ahead had been slightly burning. Then, looking at the
helium mass fraction, we see that the vortices produced a much more
mixed situation in the perturbed run, with more \emph{combustible}
fuel still present behind the flame. While the temperature is \emph{on
  average} lower (a fact visible also in the lower average height of
the fluid), a larger fraction of the fluid is burning and at deeper
depth too.  This is an important fact, especially comparing the first
and last column of \figref{fig:compa2D}, since a larger part of the
surface is hot. In terms of modelling observations, the 3D runs would
predict shorter, but brighter bursts, as can be deduced also from the
total energy release in the right panel of
\figref{fig:comparison}. This can have deep implications when trying
to extract from the observations the information about the nuclear
reactions at play. For example, using the unperturbed run profile to
fit short bursts would require much more efficient burning, changing
the conclusions about the composition, for instance, or seemingly
requiring higher yields than predicted by nuclear theory. This could
have implications for understanding the observed burst rates,
especially depending on the conclusions about the composition of the
fluid, as suggested in \citet{art-2017-cavecchi-etal}.

Looking at the curve of the maximal height in the left column of
\figref{fig:compa2D}, it is possible to identify two major ``bumps'',
one approximately at $\x = 5$ km and another one at approximately $\x
= 12$ km.  It is tempting to consider whether the 3D average profile
could be derived from the 2.5D run, since this would avoid expensive
simulations if only global values were of interest. However, apart
from the naive reduction of the time scales by a factor of $\sim 10$,
this seems a non trivial task that is beyond the scope of this paper,
and thus we will not treat it here.

Another important aspect of our results is related to the presence of
the vortices. The instability that we observed changes radically the
paradigm of flame propagation we had so far: our simulations show that
a ring of fire initiating at the equator and propagating towards the
poles will break down in vortices approximately the size of a few
Rossby radii \citep[we speculated on this
  in][]{art-2002-spit-levin-ush}. Likewise, if ignition takes place at
mid latitudes, it is likely that the flame will not propagate as an
expanding, perhaps drifting, smooth hurricane, but will instead spawn
smaller vortices after its circumference exceeds a few Rossby radii
and the flame velocities should be again $\sim 10$ times
faster. Furthermore, the local anisotropies (see in particular
\figref{fig:instabvortB}, where the height profile is shown for the
whole surface, and remember that it reflects the temperature as well)
could lead to detectable burst oscillations.

Burst oscillations are detected when performing fast Fourier
transforms of some burst lightcurves as strong spike in the frequency
domain \citep[see][]{rev-2008-gal-mun-hart-psal-chak,
  rev-2012-watts}. The mechanism producing these oscillations is still
debated, however the general idea is that the surface emission pattern
is not axially symmetric and the rotation of the star produces a
modulation of the lightcurve that can be detected in the frequency
domain. The frequency of the oscillations if often a few Hertz below
the spin of the star and it is seen increasing in the cooling phase of
the bursts. It is thought that the emission pattern must be moving
counter rotation wise on the surface of the star to produce these
effects. The most probable candidates for the mechanism of the
oscillations are a hot spot (more likely in the rising phase of the
bursts, while the flame is still propagating) or surface oscillation
modes of the burning ocean. The modelling of the pulse profiles of the
fluctuations attracts much attention, because they could be used to
measure the effects of the strong star gravity on photon propagation,
as done for the NICER detections \citep{art-2016-gend-etal,
  proc-2019-bogd}, and thus put further constraints on the neutron
star equation of state which determines the gravitational potential of
the star.

In our simulation the vortices seem to drift along the residual flow
of the fluid, which is towards the east, i.e. in the ``wrong''
direction, and, \emph{if detectable}, would lead to an observed
frequency \emph{above} the spin frequency. A first glance at the drift
in our simulation seems to indicate that there would be a
\emph{positive} difference of $\leq 1$ Hz. However, we will calculate
proper lightcurves in a followup paper. Furthermore, this simulation
describes the rising phase of the bursts, where no substantial drift
is observed \footnote{Some sources, like SAX J1808
  \citep{art-2003-chak-morgan-etal}, display a very fast chirp in the
  frequency of the burst oscillations during the rising
  phase. However, these sources display oscillations also outside the
  bursts, indicating that they possess a significant magnetic field,
  through which accretion is channeled. Having not included magnetic
  fields in these simulations a direct comparison could be
  misleading.}. When the fluid cools the flow should invert
direction. If the vortices survive up until then, they would drift in
the ``correct'' direction: when the equator is cooling, and the
pressure profile along the north direction is inverted, vortices may
drift west \citep[see][]{art-2002-spit-levin-ush}. Also, even if the
vortices themselves are not the direct origin of the burst
oscillations in the tail of the bursts, they could well be what
\emph{excites} the specific surface modes that produce the
oscillations.

Finally, in the case of mid-latitude ignition, the breaking up of
the expanding hurricane into sub-systems may weaken its signature as
burst oscillations and this would explain why they are so difficult to
detect during the rise of the bursts. The expected launch of new
telescopes like \textit{eXTP} \citep{art-2016-zhang-etal} and
\textit{STROBE-X} \citep{art-2017-wilhod-etal} present an
unprecedented opportunity. With their high collecting area, which
allows for fast time resolution also during the rising part of the
lightcurves, they may be able to detect the traces of these structures
in future observations of the type I bursts and shed further light on
the mechanism of the burst oscillations.

\emph{Acknowledgements: } YC is supported by the European Union
Horizon 2020 research and innovation programme under the Marie
Sklodowska-Curie Global Fellowship grant agreement No 703916. YC would
like to thank Mike Zingale for useful discussions. This simulations
ran on the Perseus cluster at Princeton University and we would like
to thank IT for their support.

%thebibliography
\bibliographystyle{aasjournal}
\bibliography{ms}

\appdo{}

\label{lastpage}

%end of document
\end{document}